\documentclass[]{article}

\usepackage[centertags]{amsmath}
\usepackage{amssymb}
\usepackage{theorem}
\usepackage{latexsym}
\usepackage{times}
\usepackage{ifthen}
\usepackage{xspace}
\usepackage[dvips]{graphicx}
\usepackage{subfigure}

\graphicspath{{$PWD/}{/homes/mcguire/net3/}}
\newcommand{\R}[1][1]{\ensuremath{\Rset\ifthenelse{\equal{#1}{1}}{}{^{#1}}}\xspace}
\newcommand{\C}[1][1]{\ensuremath{\mathbf{C}\ifthenelse{\equal{#1}{1}}{}{^{#1}}}\xspace}
\newcommand{\N}[1][1]{\ensuremath{\mathbf{N}\ifthenelse{\equal{#1}{1}}{}{^{#1}}}\xspace}

\newcommand{\eps}{\ensuremath{\varepsilon}\xspace}

\renewcommand{\vec}[1]{\ensuremath{{\boldsymbol{#1}}}\xspace}

\begin{document}

\title{Threshold Disorder as a Source of Diverse and \\
                   Complex Behavior in Random Nets}

\author{Patrick C. McGuire
\footnote{Department of Physics;
          University of Arizona;
          Tucson, AZ 85721, USA}\, 
\footnote{Neuroinformatics Group: Computer Science Department:
   Technische Fakult\"at (Engineering College); University of Bielefeld; 33501 Bielefeld, Germany}\,
\footnote{Center for Interdisciplinary Studies (ZiF): Complexity Program;
       University of Bielefeld, 33501 Bielefeld, Germany}\,
\footnote{Before March 15, 2002, please correspond with this author at his
   current address in the University of Bielefeld Neuroinformatics Group;\break
        or by Phone (0049) 521-106-6059 or -6060; FAX (0049) 521-106-6011;
          or at his email address: mcguire@techfak.uni-bielefeld.de}\,
\footnote{After March 15, 2002, please correspond with this author at his
   next address in the
   Centre de Astrobiolog\'{\i}a (CSIC/INTA); Instituto Nacional de T\'ecnica
   Aeroespacial; Ctra de Torrej\'on a Ajalvir, km 4;
   28850 Torrej\'on de Ardoz; Madrid, Spain; or by Phone (0034) 91-520-21-07;
   FAX (0034) 91-520-10-74, or at his email address:
   mcguire@cab.inta.es}\, ,
Henrik Bohr
\footnote{Department of Physics; DTU, The Technical University of Denmark,
          B. 307; DK-2800 Lyngby, Denmark}\,
\footnotemark[1]\,
\footnotemark[3]\, ,
John W. Clark
\footnote{Department of Physics and the McDonnell Center
          for the Space Sciences; Washington University; St. Louis,
          MO 63130, USA}\,\,\, 
\footnotemark[3]\, ,
\\
Robert Haschke
\footnotemark[2]\, ,
Chris L. Pershing
\footnotemark[1]\,
\footnote{Biomedical Engineering Program;
          University of Arizona; Tucson AZ 85721, USA}\,\,\, ,
Johann Rafelski\footnotemark[1]\\
}
\date{\today}
\maketitle

\begin{abstract}
 We study the diversity of complex spatio-temporal patterns in the
behavior of random synchronous asymmetric neural networks (RSANNs).
Special attention is given to the impact of disordered threshold
values on limit-cycle diversity and limit-cycle complexity in RSANNs
which have `normal' thresholds by default.  Surprisingly,
RSANNs exhibit only a small repertoire of rather complex limit-cycle
patterns when all parameters are fixed. This repertoire of complex
patterns is also rather stable with respect to small parameter changes.
These two unexpected results may generalize to the study of other
complex systems. In order to reach beyond this seemingly-disabling
`stable and small' aspect of the limit-cycle repertoire of RSANNs,
we have found that if an RSANN has threshold disorder above a critical
level, then there is a rapid increase of the size of 
the repertoire of patterns. The repertoire size initially follows
a power-law function of the magnitude of the threshold disorder.
As the disorder increases further,
the limit-cycle patterns themselves become simpler until at a second
critical level most of the limit cycles become simple fixed points.
Nonetheless, for moderate changes in the threshold parameters, RSANNs
are found to display specific features of behavior desired for
rapidly-responding processing systems: accessibility to a large set of
complex patterns. 

\end{abstract} 

\pagebreak

\section{Introduction}
Random Synchronous Asymmetric Neural Networks (RSANNs) with fixed
synaptic coupling strengths and fixed neuronal thresholds/inputs tend 
to have access to a very limited set of different limit cycles
(Amari (1974), Clark, K\"urten \& Rafelski (1988),
Littlewort, Clark \& Rafelski (1988), Hasan (1989),
Rand, Cohen \& Holmes (1988), Clark (1990), Schreckenberg (1992)).
We will show here, however, that when we add a moderate amount of randomly
quenched noise or disorder, by choosing the neural thresholds or inputs
to vary within a prescribed gaussian distribution, we can gain controllable,
and we believe biologically relevant, access to a wide variety of
limit cycles, each displaying dynamical participation by many neurons.

  The appearance of limit-cycle behavior in central pattern generators
is evidence for cyclic temporal behavior in biological
systems (Hasan (1989), Marder \& Hooper (1985)). Previous computational
models, as discussed above, do not exhibit a diverse repertoire of
limit-cycle behaviors, as biological systems often demonstrate (e.g.,
the different gaits of a horse, or the different rhythmic steps
of a good human dancer).  Additonally, it is our belief that the
biologically-interesting networks are those in which a significant
fraction of the neurons can (and often do) participate in the local
dynamics. In principle, spatially-sparse neuronal firing patterns can
be constructed from a large network of strongly-participatory neurons
by self-organized architectural inhibition of selected
neuronal assemblies. This can leave the uninhibited
neuronal assemblies able to freely participate in the neural
dynamics (for some time), though these uninhibited neurons or
neuronal assemblies may be isolated in space from each other, only
connected to other active neurons through non-local or indirect
connections (Gray \& Singer (1990)). Therefore, in this paper, we explore
the problem of how to produce a computational neural model
which possesses a diverse repertoire
of strongly-participatory limit-cycle behaviors.

A system which can access {\it many} limit cycles should always be
able to access a novel mode; hence the system would have the potential
to be a `{\it creative}' system. Herein, we demonstrate
conditions sufficient to allow a simple computational neural system
to access creative dynamical behavior.

In Section~\ref{sec:RSANN} we introduce RSANNs along with the concept
of threshold disorder, as well as a measure to distinguish different
limit cycles. In our quantitative investigations we need to introduce,
with some precision, concepts which intuitively are easy to grasp, but
which mathematically are somewhat difficult to quantify. We define
`{\sl eligibility}' in Section \ref{sec:Elig} as an entropy-like measure
of the fraction of neurons which actively participate in the dynamics
of a limit cycle. In order to quantify the RSANN's accessibility to
multiple limit-cycle attractors, we define `{\sl diversity}' in
Section~\ref{sec:Div} as another entropy-like measure, calculated from
the probabilities that the RSANN converges to each of the different
limit cycles.  The difference between eligibility and diversity is
that the former applies to a limit cycle observed in a specific
network, while the latter applies to the collection of limit cycles
that the network can exhibit. To measure the creative potential
of a system, we introduce the concept of `{\sl volatility}' as the
ability to access a huge number of highly-eligible cyclic modes.
We find that in terms of these
diagnostic variables, as the neuronal threshold disorder $\eps$
increases, our RSANN exhibits a phase transformation at $\eps=\eps_1$
from a small number to a large number of different {\em accessible}
limit-cycle attractors (Section~\ref{sec:Div}), and another phase
transformation at $\eps=\eps_2 > \eps_1$ from high eligibility to low
eligibility (Section~\ref{sec:Elig}).  Our main result is that the
volatility is high only in the presence of threshold disorder of suitably
chosen strength between $\eps_1 \leq \eps \leq \eps_2$, thereby 
allowing access to a diversity of eligible limit-cycle attractors
(Section~\ref{sec:Vol}).

\section{Random asymmetric neural networks with threshold or input noise}
\label{sec:RSANN}
Symmetric neural networks (SNNs) (Hopfield (1982)) became widely used
in associative memory applications due to their ability to store a
large number of patterns as fixed points of their dynamics; however,
their dynamical behaviour is restricted to fixed points or 
limit cycles of period 2. In contrast, asymmetric neural networks (RSANNs)
show a complicated dynamical behaviour, including limit cycles
of large periods or even chaos
\footnote{for networks of binary-valued neurons,
the dynamical behavior can simulate chaos,
but for networks of real-valued neurons, true
chaos is observable, with the prerequisite for chaos:
 `sensitive dependence upon initial conditions'.}
(Amari (1974), Clark, Rafelski \& Winston (1985),
Clark, K\"{u}rten \& Rafelski (1988), Littlewort, Clark \& Rafelski (1988),
K\"{u}rten (1988), Bressloff \& Taylor (1989),
Clark (1990, 1991), McGuire, Littlewort \& Rafelski (1991),
McGuire {\it et al.} (1992),
Bastolla \& Parisi (1997)).
Moreover, they offer considerably more biological realism, since real neuronal
 connections tend to be unidirectional.

We investigate a network of $N$ threshold elements, i.e. their firing
states have binary values $x_i \in \{0, 1\}$. Each neuron $i$ is connected
to $M < N$ presynaptic neurons by unidirectional weights
$w_{ij}$, with $w_{ij} \ne w_{ji}$ and $w_{ii} = 0$.
All weights are independent random variables, drawn from a
uniform distribution within $[-1, 1]$. A neuron fires if its
post-synaptic-potential (PSP) is greater than its specific threshold
$V_i$. Therefore the network is described by the following system of
equations for `sum-and-fire' McCullough-Pitts neurons:
\begin{align}
  x_i (t+1) &= \Theta \bigl(\sum_{j=1}^M w_{ij} x_j(t) - V_i\bigr)
  \quad \forall i \in 1, \dots, N \, ,
  \label{eq:Dynamics}
\end{align}
where $\Theta$ is the Heaviside function. Supposing that all neurons should
actively participate in the dynamics, with a mean firing rate
$\langle x_i \rangle = \frac{1}{2}$, the mean thresholds $V^0_i$ are
adjusted so that the mean overall input
\begin{align} \langle u_i \rangle =
\langle \sum w_{ij} x_j - V^0_i \rangle \approx \sum w_{ij} \langle x
\rangle - V^0_i\,
\label{eq:approx}
\end{align}
to a generic neuron $i$ becomes zero
(so that it is poised on the boundary between firing
and not firing). Thus, we have:

\begin{align}
  V^0_i &= \frac{1}{2} \sum_j w_{ij} , \\
  V_i &= \eta_i V_i^0 , 
  \label{eq:ThreshNoise1}
\end{align}
where the parameter $\eta_i$ is chosen to modulate the threshold.
The case $\eta_i \equiv 1$ for all $i$ corresponds to the choice known as
`normal' thresholds (Clark (1991)).  The mean firing
rate of 0.5 is quite high biologically,
but computationally, it is a reasonable point to begin our research;
it is not too difficult to adapt the treatment to
lower mean firing rates. In order
for a given amount of threshold disorder to affect all neurons more-or-less
equally, we have chosen
here a multiplicative scaling of the thresholds relative to the normal
thresholds rather than an additive scaling. We do not consider
synaptic noise or modulation; hence the weights $w_{ij}$ are 
kept fixed for a given network.

In considering the living neural networks in the brain, some
researchers treat the neuronal thresholds as constant and
noiseless (as in the Hodgkin-Huxley and Fitzhugh-Nagumo models;
see Murray (1989) for a summary); others are convinced that
neurons live in a very noisy environment, both chemically and
electrically, with nontrivial consequences for neuronal and
network function (see Zador (1997), Chow \& White (1996),
Clark (1988), Buhmann \& Schulten (1987), Shaw \& Vasudevan (1974),
Little (1974), Taylor (1972), and Lecar \& Nossal (1971)).
Examining the issue more closely, we may note that Mainen \& Sejnowski
(1995) have presented data suggesting a low intrinsic noise level for
neurons, which does not seriously affect the precision of spike timing
in the case of stimuli with fluctations resembling synaptic activity.
On the other hand, Pei, Wilkens \& Moss (1996) have presented evidence
that noise can exert beneficial effects on neural processing through
the phenomenon of stochastic resonance.

\begin{figure}[h]
  \centering
  \includegraphics[width=13.0cm]{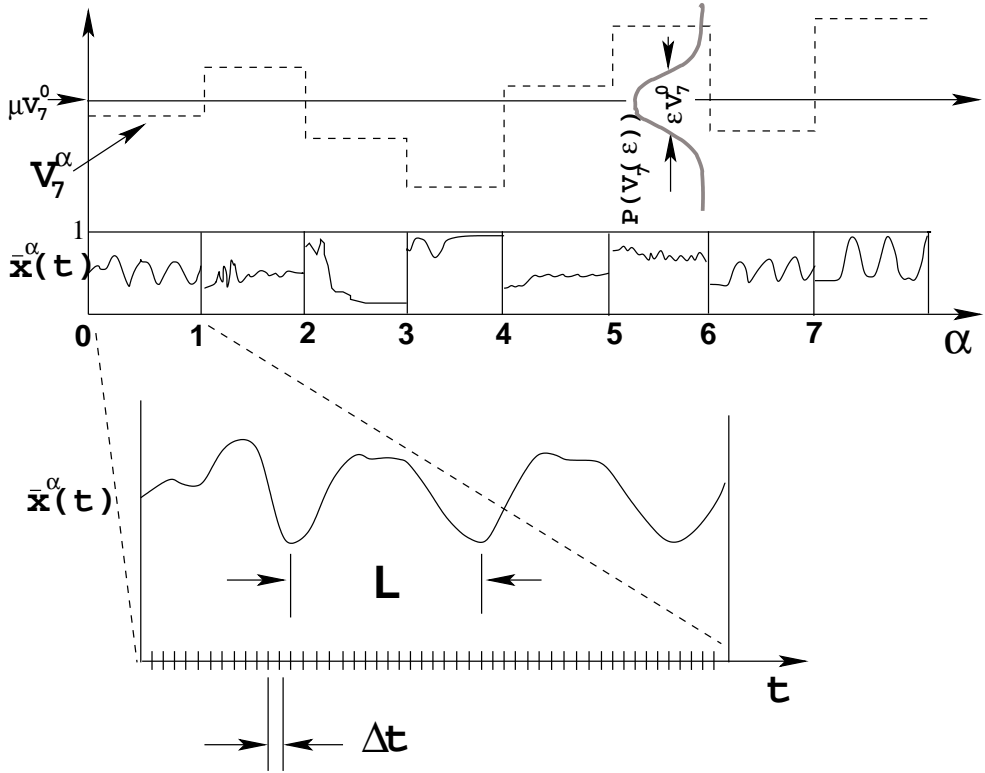}
  \caption{ Qualitative sketch showing the varying threshold
    for neuron \#7 ($V_7^\alpha$) and the spatially-averaged firing
    rate $\bar{x}^\alpha(t)$, as a function of trial-number $\alpha$,
    and in the inset, a magnified view of $\bar{x}^\alpha(t)$, all as
    functions of time, $t$.}
  \label{fig:clock}
\end{figure}

  Mathematically, the external inputs to a neuron from sensory organs or from
other areas of the neural system can also be treated as a modulation
of the threshold of that neuron. This suggests that the
results obtained on threshold modulation might be
easily generalized to the situation of external modulation.

Taken together, the noisiness of thresholds and the variability of
inputs can be viewed as a changing environment. 
We simply model this complex
changing environment by varying the normal thresholds
using multiplicative gaussian noise $\eta_i$ with mean $\mu=1$ and
standard deviation $\eps$, leading to Eq.~\ref{eq:ThreshNoise1}. The
components $\eta_i$ are chosen independently for all neurons $i$. It
may be much more reasonable to consider spatially-correlated noise
amplitudes, but such a study exceeds the scope of our present effort.

\subsection{Limit-Cycle Search}
\label{sec:Updating}

Since we wish to study the diversity of different limit
cycles accessible with small changes of the thresholds, 
we need a robust criterion for detecting limit cycles.
Even in the presence of small-amplitude noise effective on a shorter
time scale than the cycle
length, the neural net will never stabilize into a detectable perfect
cycle. Rather, it will either converge to an {\it approximate} limit cycle with
occasional misfirings or never converge at all.
Such approximate limit-cycle behavior is more relevant to neurobiological
systems than is its perfect realization, due to the inherent destabilizing
noise (from membrane-potential or synaptic noise) and additional
complicating factors, notably (1) the complexity of biological
neurons, (2) the continuum of signal transmission times between
neurons, and (3) the apparent lack of a clock to synchronously update
all neurons.

However, although approximate limit-cycle behavior might be more common in
volatile systems, it is not ideal for computer simulation and
computer characterization. Therefore, for the sake of the computational
tractability, we restrict our search to perfect limit cycles.
In order to achieve this, we fix
the neural thresholds $V_i$ until a limit cycle is found during
network evolution via Eq.~\ref{eq:Dynamics}. Since the
noise is frozen-in (`quenched')
for a long period, it is more properly considered as 
disorder.  Before the next trial, the thresholds are varied
according to Eq.~\ref{eq:ThreshNoise1}, changing the underlying
network dynamics; they are then fixed again during limit cycle search.
Each trial step starts with the activity pattern $\vec{x}(0)$ with
which the previous trial was terminated. To gain a qualitative
understanding of our approach, see Figure \ref{fig:clock}.

Fixed points (with limit-cycle period $L = 1$) are generally of less interest
than cyclic modes in view of the spontaneous oscillatory behavior
displayed by real neural systems (e.g walking or singing).
Effectively chaotic or non-cycling
behavior (with $L \sim 2^N$) is not predictable enough to be of much
use for most applications in real neural systems.  Limit cycles of
intermediate period are consequently our paramount concern.
 
We update all neuron firing states in parallel, or `synchronously', as
opposed to either serial or random updating in which only one
neuron is updated at a given time step.

In our simulations, we used $M = N/10$ incoming connections
per neuron, where self-connections were not allowed, and we studied
networks with $N \in \{10,20,30,40,50,100\}$ neurons. Self-connections
tend to have a stabilizing effect on the network, often driving
the behavior towards a fixed point with only very brief transients.
The different behavior of networks with and without self-connections
might be a worthy subject
of future investigation, but we chose not to emphasize that direction here.
For practical computational reasons, the network sizes investigated
are primarily constrained by the existence of
extremely long limit cycles and transients
of large networks, occurring especially when the thresholds are near-normal,
(see Clark (1990, 1991), Clark, K\"urten \& Rafelski (1988),
 and Littlewort, Clark \& Rafelski (1988)
for discussions of normal thresholds and the
correlation between transient length and limit-cycle period). A
network of $N$ threshold units can assume $2^N$ states, placing an 
upper limit on the length of a limit cycle. This upper limit 
for the cycle length is due to the facts that there are only a finite number
of states and that the time development of the system is
deterministic and depends only on the initial network state $\vec{x}(0)$.

Since the detection of limit cycles at the microstate level
$\vec{x}$ is too time consuming (it requires ${\cal O}(N L^2)$
comparisons), we use the system-averaged firing rate $\bar{x}(t)$
in order to test for periodicity:
\begin{align}
  \label{eq:Activity} 
  \bar{x}(t) &\equiv \frac{1}{N} \sum_{i=1}^{N} x_i(t) \, , \\ 
  \label{eq:Activity2} 
  \bar{x}(t+L) &= \bar{x}(t) \quad \forall t \in [0, 4 L] \, ,
\end{align}
where the limit-cycle period is is identified as $L$. Though satisfying
Equation~\ref{eq:Activity2} is only a necessary condition
for an exact limit cycle of period $L$
at the microstate-level, in practice we observed no differences
between exact and average comparison methods on a small test set.
We used a window of $4 L$ time steps to ensure that
the cycle does indeed repeat itself in $\bar{x}$ four times;
without explicitly tracking the microstate $\vec{x}$, such
care is necessary in order to avoid false limit-cycle
detection and measurement.

\subsection{Limit-cycle Comparison}
Since diversity and volatility (which we define in
sections~\ref{sec:Div}~\&~\ref{sec:Vol})
require an abundance of {\em different} limit cycles, we need to
introduce a high-contrast, direct, neuron-by-neuron measure to decide
whether a given limit cycle is different from or similar to another
limit cycle. One could, of course, just compare the full neuron-by-neuron
time-dependence of the activity patterns of the cycles themselves,
but that would
require a vast amount of memory to store all observed cycles. 
However, if as above, we choose to compare the time-dependence of the
system-averaged firing rates instead of the full firing vectors,
different limit cycles may be remarkably similar, 
possibly distinguished by only a small numerical difference, which
we elucidate here with a specific example.
Given that:
\begin{itemize}
\item a network of $N$ neurons has two cycles:
cycle {\bf A} with period $L$ and cycle {\bf B} with period $2 L$,
\item cycle {\bf A} has the same firing pattern at each time step
as cycle {\bf B} with the exception of two neurons at each time step
in cycle {\bf A} which differ from the corresponding two neurons in cycle {\bf B},
\item and the total number of firing neurons at each time step in cycle {\bf A}
is the same at each time step in cycle {\bf B} (due to the two neurons
cancelling each other), 
\end{itemize}
then with  system-averaged firing rates, the two cycles would be deemed
identical, with the same period; though a comparison of the time-dependence of
full firing vectors would show the different period of the two cycles. 
 
Reliable discrimination clearly requires a compact
measure, i.e. a fingerprint of a cycle, which is:
\begin{itemize}
\item independent of cycle length,
\item capable of discriminating between a wide array 
of limit cycles, and
\item easily computable.
\end{itemize}
 For this purpose we use
the vector $\bar{\vec{x}} = [\bar{x}_1, \dots, \bar{x}_N]^t$ formed by the 
time-averaged firing rates within $T=4 L$ time steps, 
\begin{equation}
  \bar{x}_i = \frac{1}{T} \sum_{t=1}^{T} x_i(t).
  \label{eq:TimeAvFiringRate}
\end{equation}
Since we base our limit-cycle detection upon a temporal
quantity (the time-dependent, system-averaged firing rate),
our additional reference to a `spatial' quantity (the time-independent,
time-averaged firing vector) serves as a good cross-check.
Obviously, use of limit-cycle period alone as our measure
of similarity might have commonly led to misclassified cycles.

\begin{figure}[p]
  \centering
  \begin{minipage}{1.25\textwidth}
  \subfigure[cycles with equal periods]{%
    \includegraphics[width=0.28\textwidth]{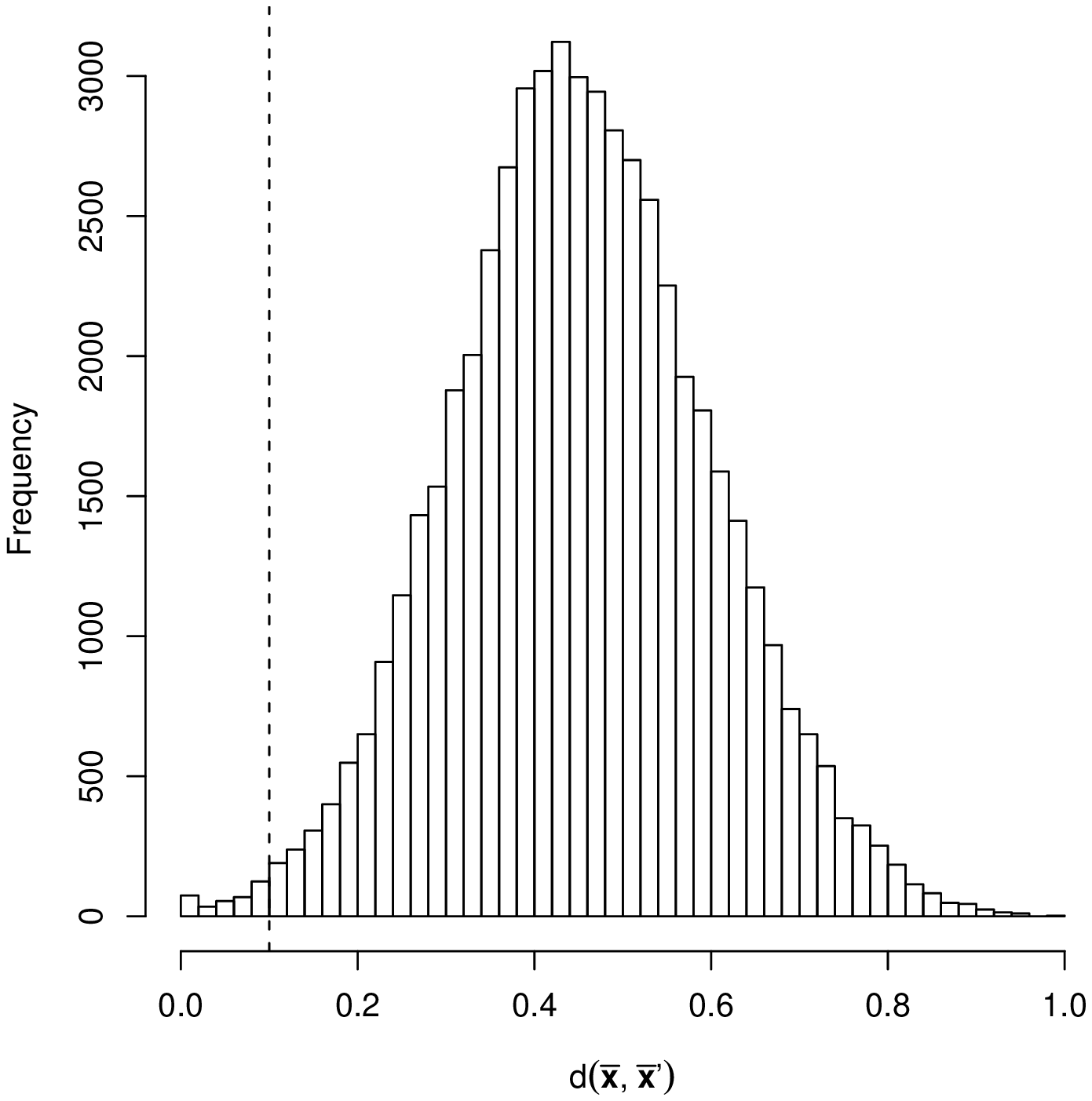}
    \includegraphics[width=0.28\textwidth]{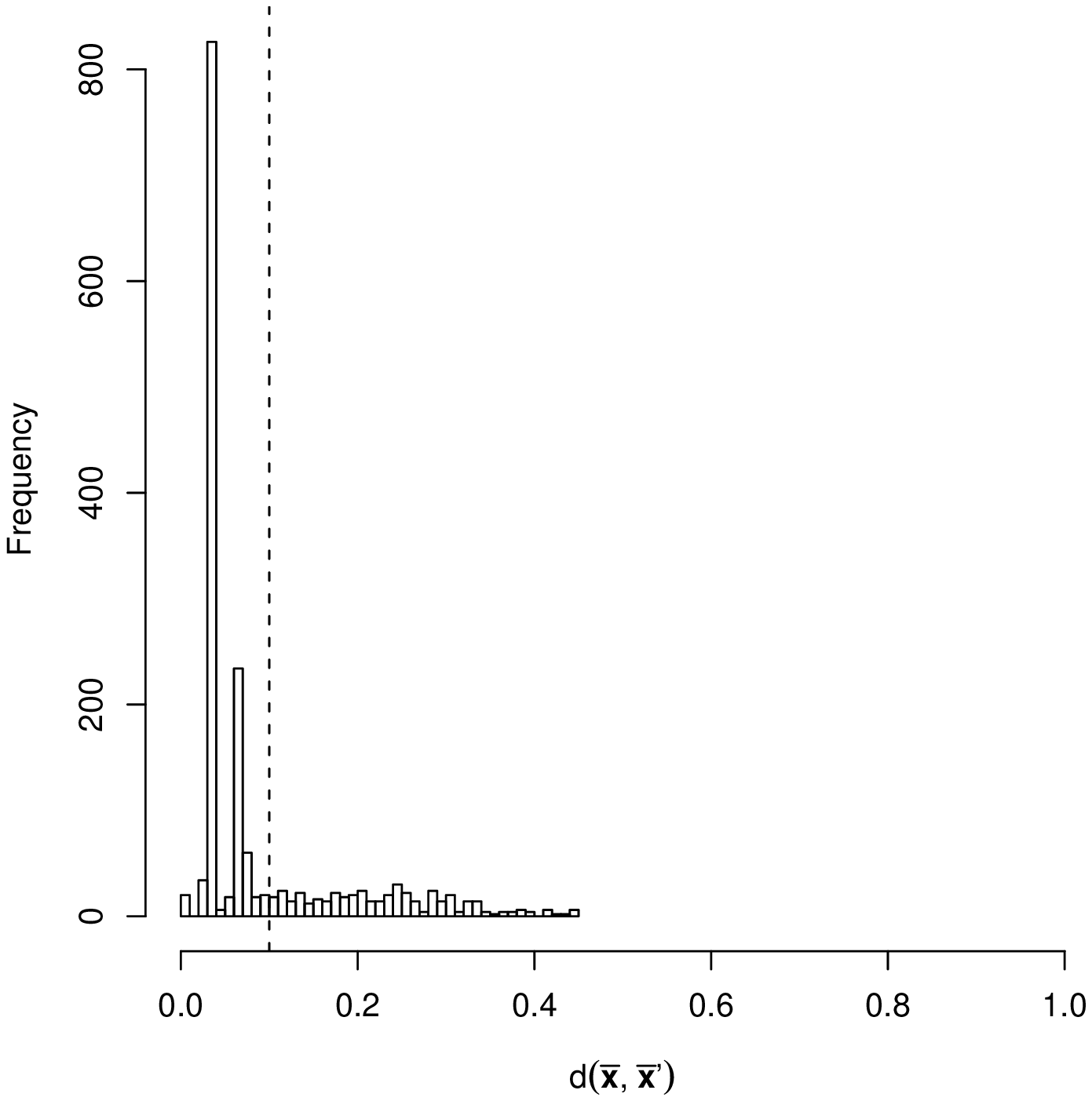}
    \includegraphics[width=0.28\textwidth]{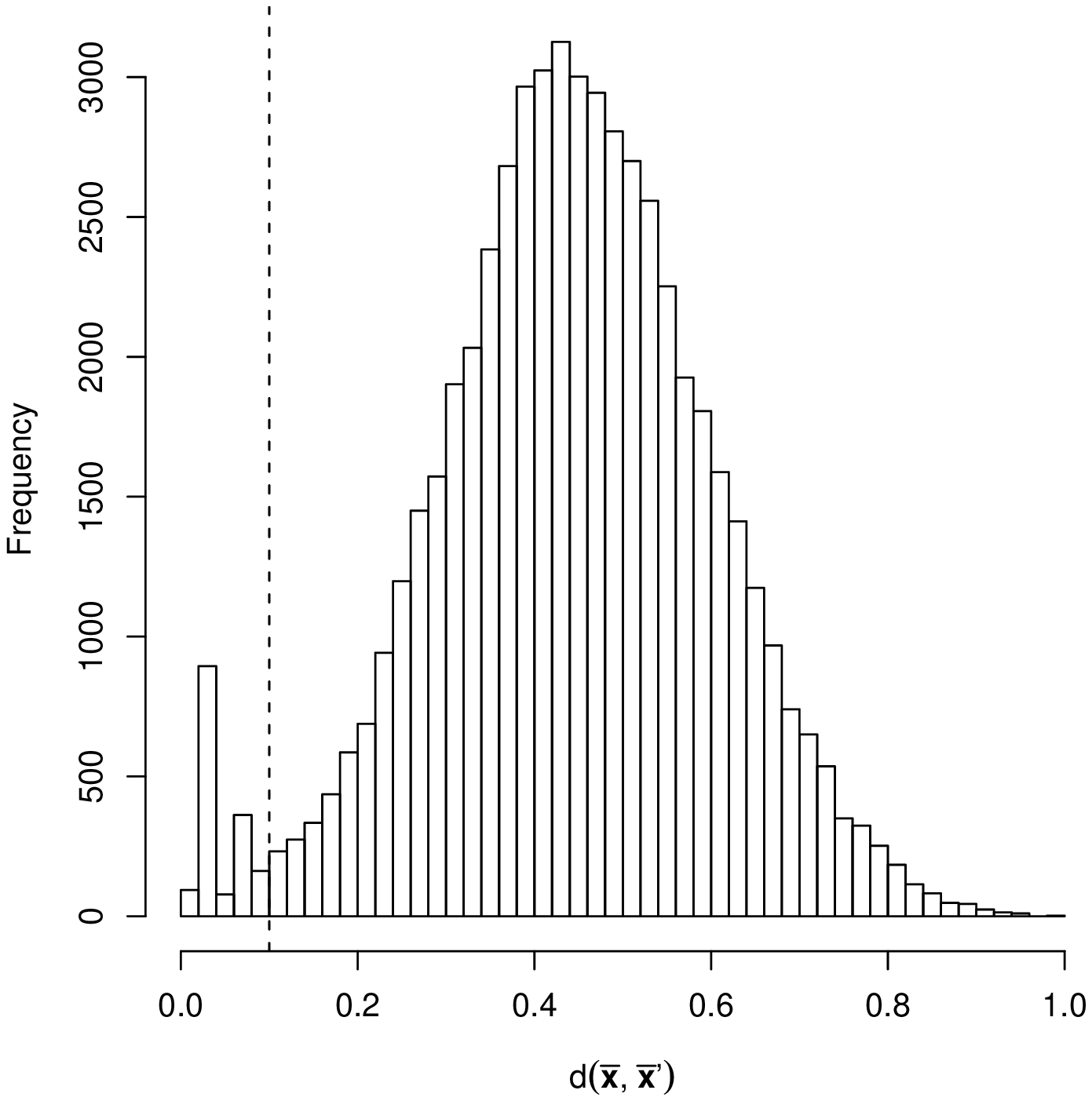}}
  \subfigure[cycles with different periods]{%
    \includegraphics[width=0.28\textwidth]{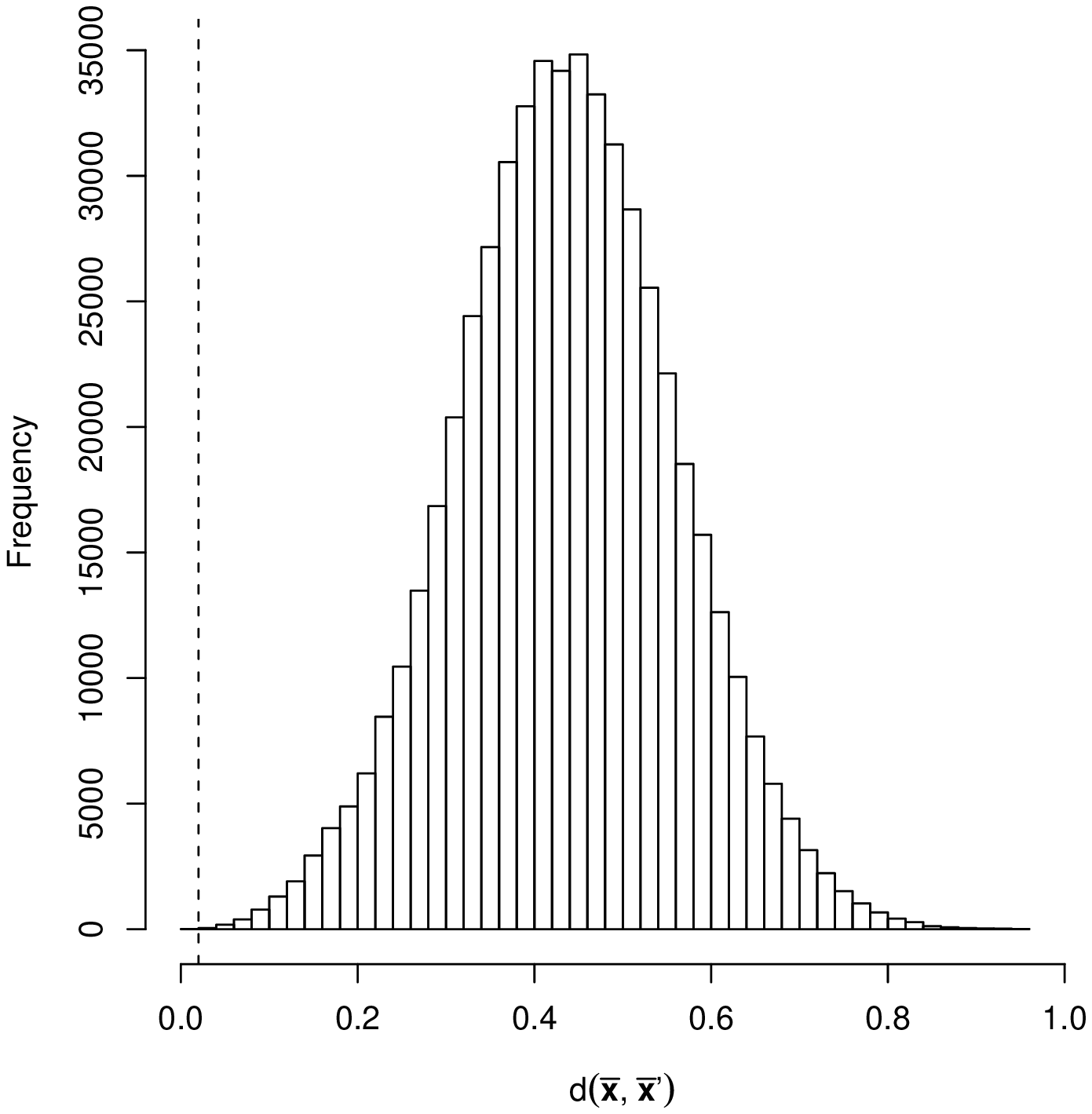}
    \includegraphics[width=0.28\textwidth]{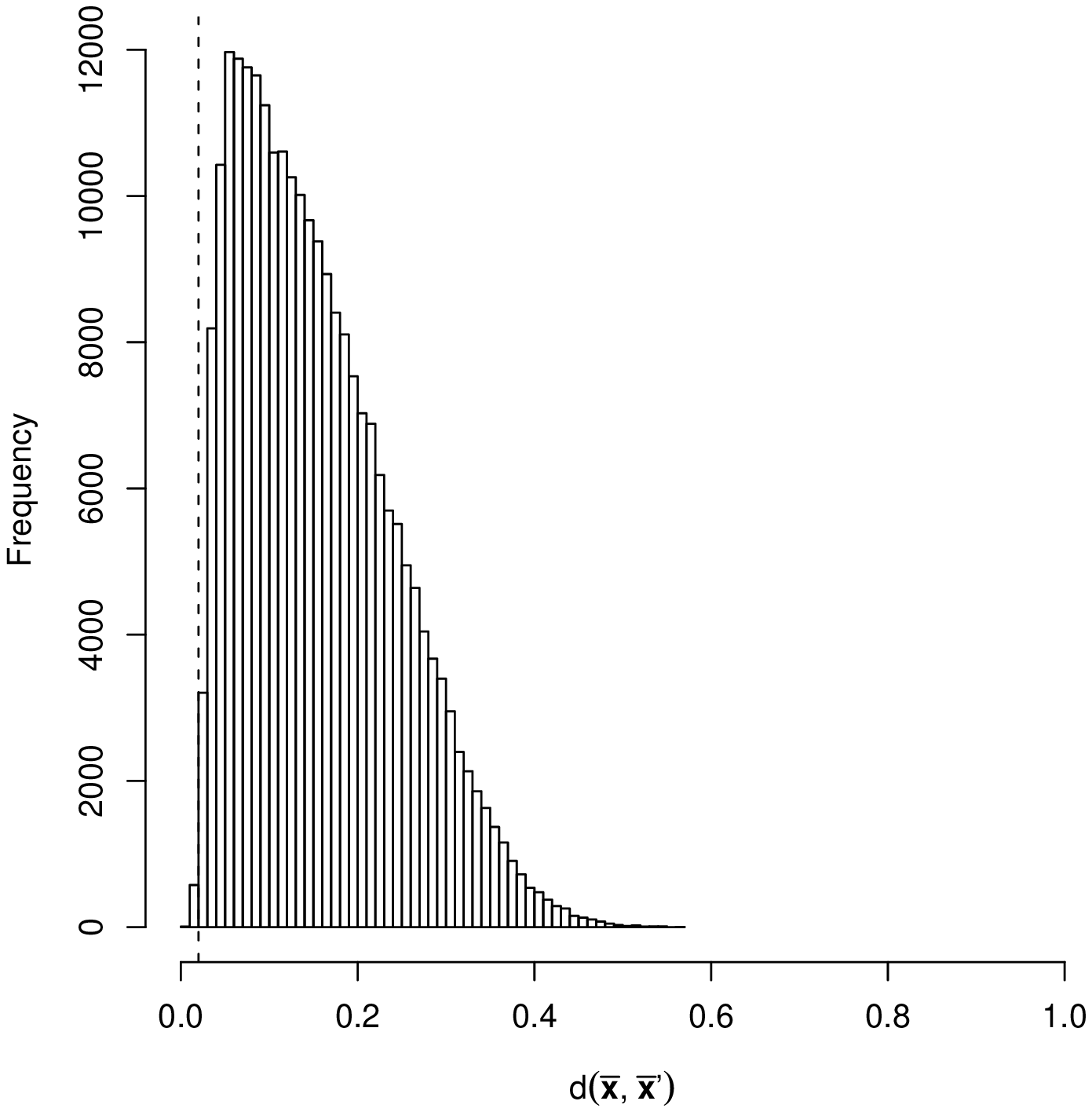}
    \includegraphics[width=0.28\textwidth]{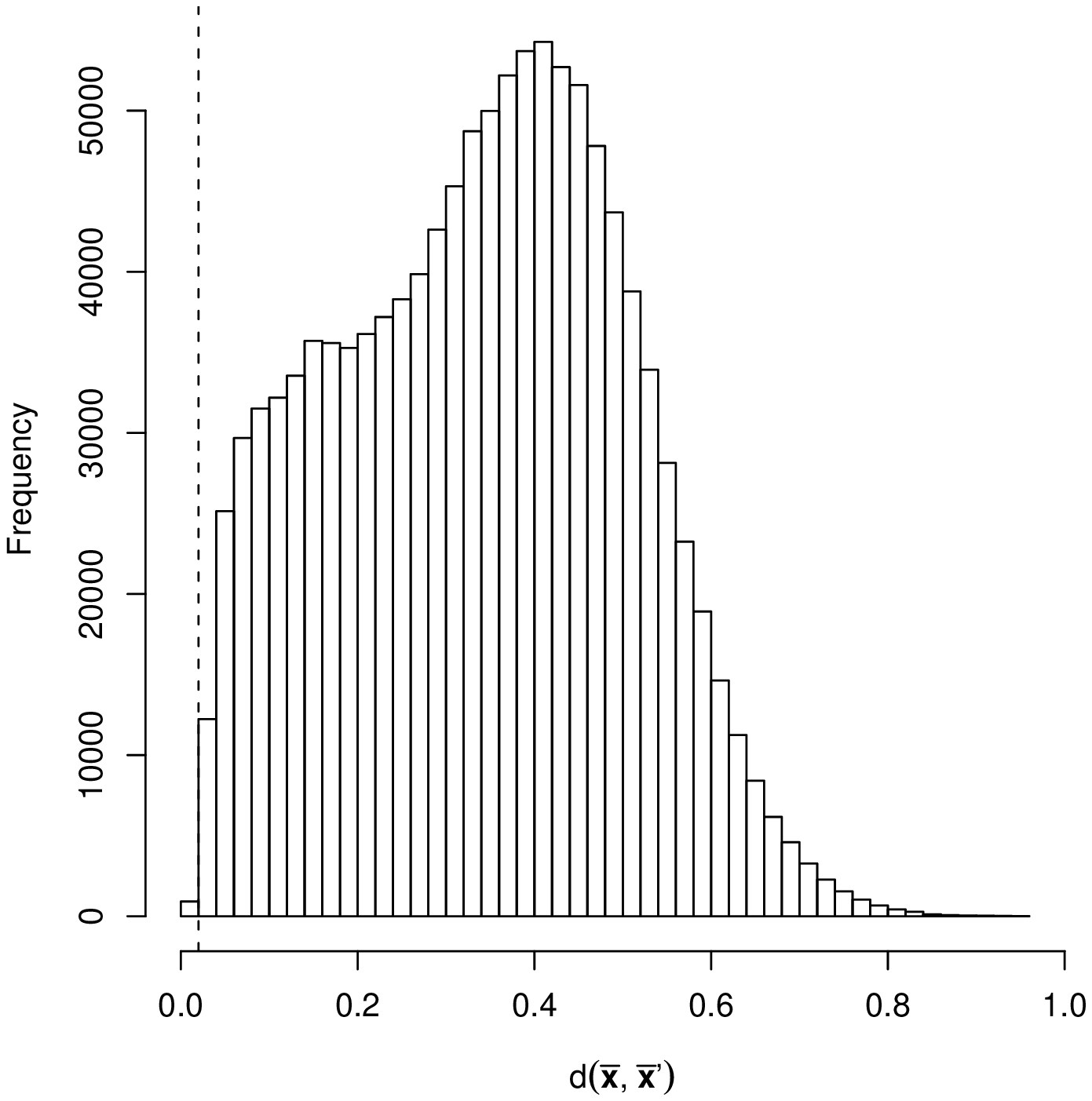}}
  \subfigure[all cycles]{%
    \includegraphics[width=0.28\textwidth]{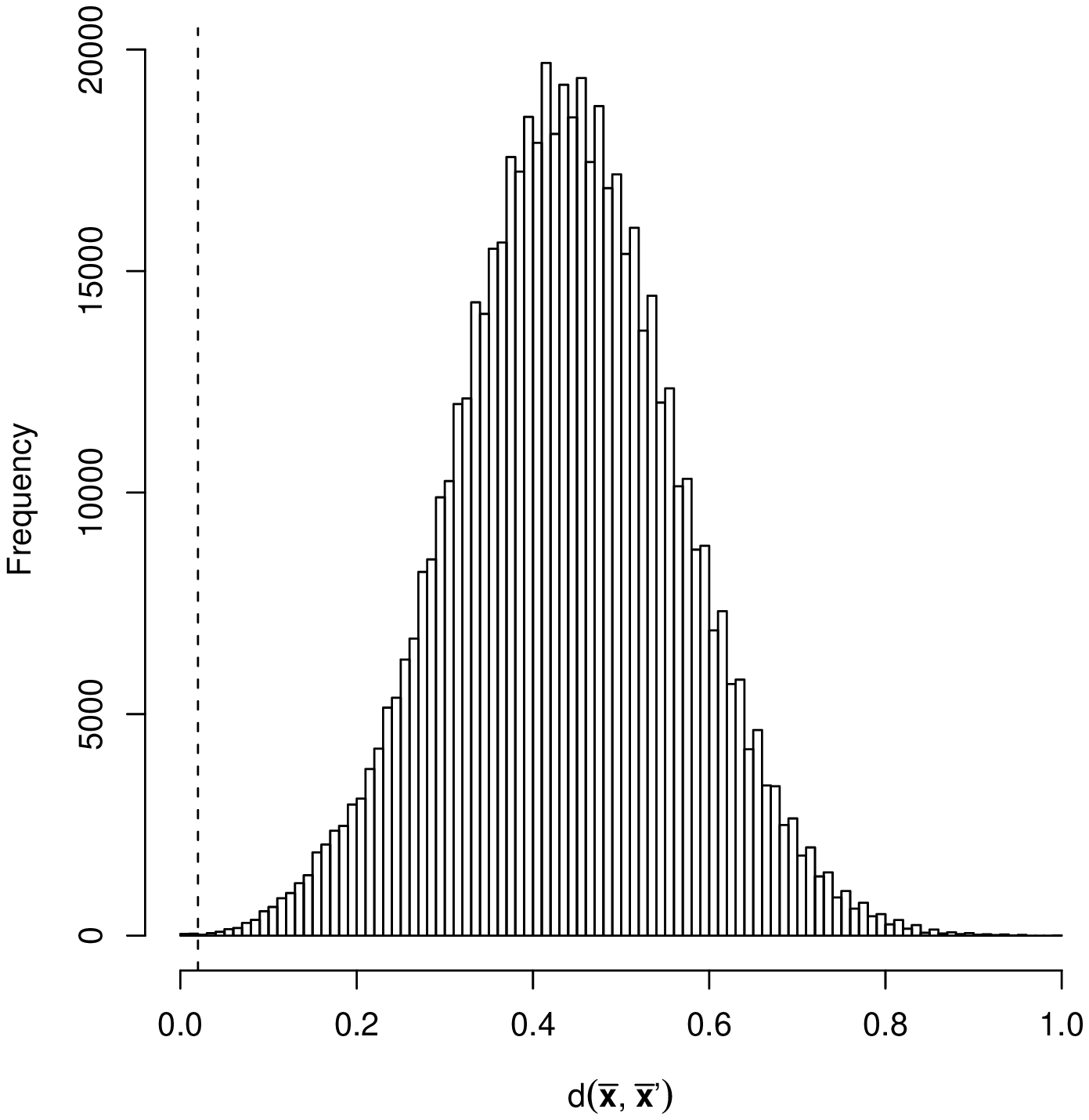}
    \includegraphics[width=0.28\textwidth]{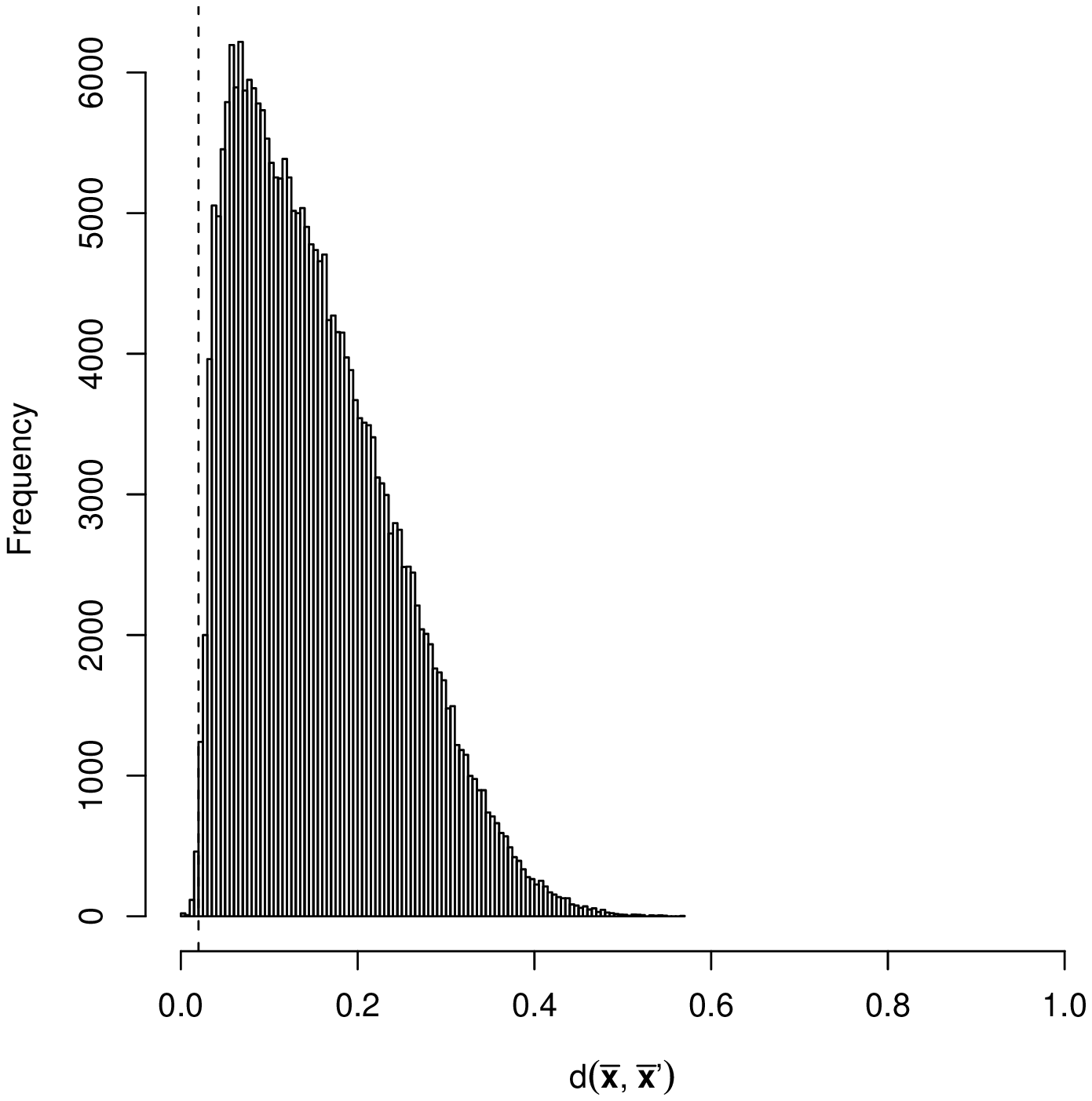}
    \includegraphics[width=0.28\textwidth]{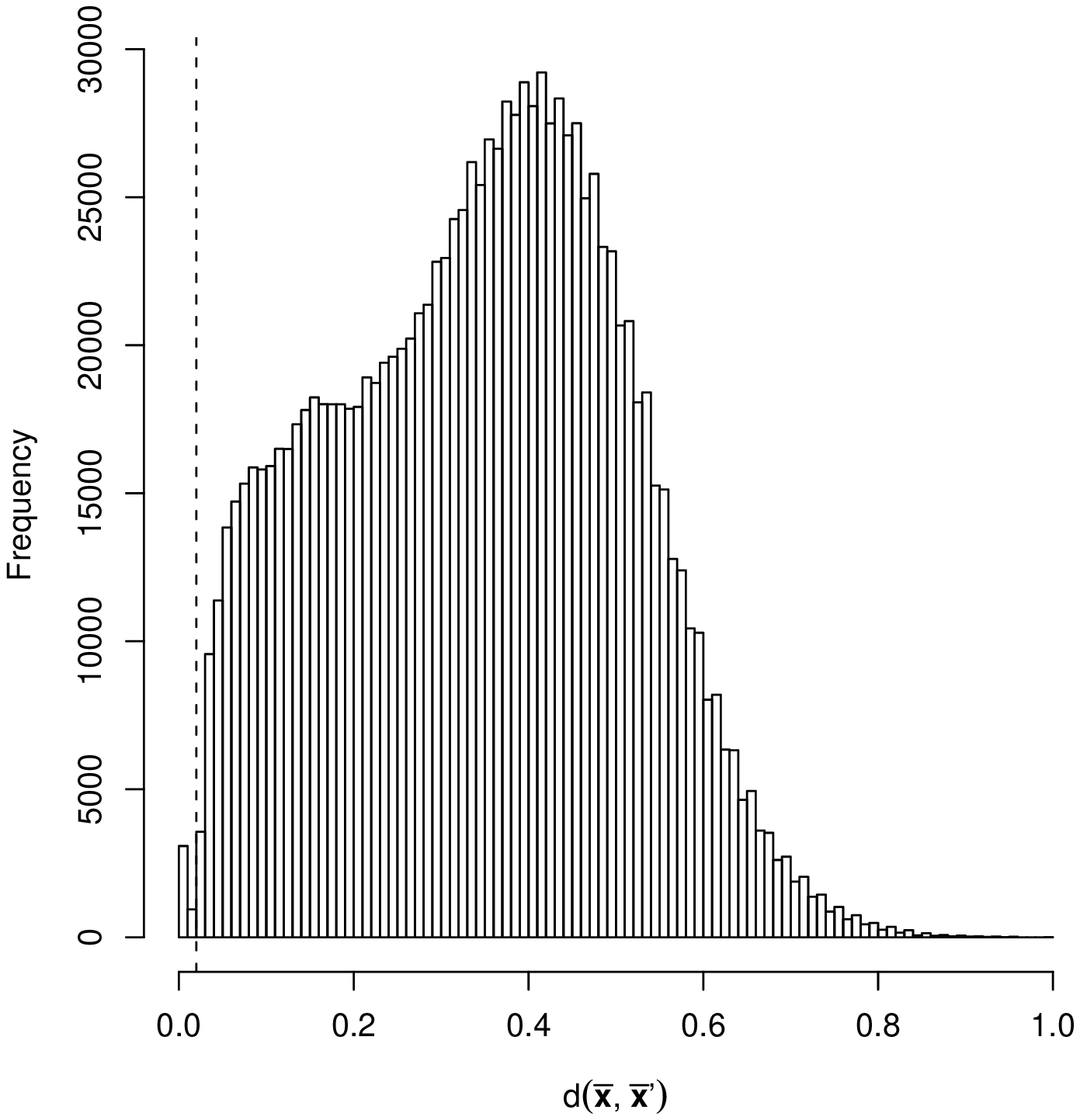}}
  \end{minipage}
  \caption{Histogram of cycle distances $d(\bar{\vec{x}},\bar{\vec{x}}')$
    for small periods ($L < 50$) (left column), large periods ($L >
    50$) (center column), and arbitrary periods (right column).
    Based partly upon this figure, the threshold for regarding a cycle pair
    as similar or different is chosen to be $d=0.02$. In the text, we
    discuss the exception of $d=0.1$ for cycle-pairs with equal periods.}
  \label{fig:DiffHisto}
\end{figure}

We do not want to distinguish between very similar limit
cycles, separated by only a small number of misfirings. Therefore,
for two cycles to be considered similar, we
allow small non-zero values of the distance 
\begin{equation}
  d(\bar{\vec{x}},\bar{\vec{x}}') = 
  \frac{1}{N} \sum_{i=1}^{N} |\bar{x}_i - \bar{x}_i'|
\end{equation}
between their fingerprints $\bar{\vec{x}}$ and $\bar{\vec{x}}'$.

In Figure~\ref{fig:DiffHisto}, we show histograms of the distances
between limit cycles with (a) equal and (b) different lengths,
accumulated over a large number of different cycles and all investigated
networks and thresholds.  Due to their high frequency, we have
excluded pairs of cycles with $d=0.0$ from this figure.  If the two
cycles have the same period $L$, the overall number of misfirings
during $T$ timesteps is given by $N T d$.  The chosen value of the threshold
$d_{\text{max}}=0.02$ for cycles to be regarded as similar is
indicated in Figure~\ref{fig:DiffHisto} as well. Evidently, this
choice is low enough to prevent misclassification of most of the
different limit cycles, as it avoids the large peaks for larger $d$,
and it is sufficiently larger than zero to tolerate small deviations
in similar cycles. It corresponds to one `misfiring' per time step
for $N=50$, on average. 

As observed in Figure~\ref{fig:DiffHisto}.a (center column), cycle
pairs with equal and long periods ($L>50$) produce a curious
clustering of distances within $0.02< d < 0.1$. A closer inspection of this
peak reveals that only three cycle pairs contribute to this clustering.
Therefore, in order to be conservative, we raised
the threshold for limit-cycle difference to $d_{\text{max}} = 0.1$ in
this case. Since the number of cycle pairs within this peak is only a
small fraction of the total number of cycle pairs observed, this
change in  $d_{\text{max}}$ changes the `diversity' and `volatility' (defined
later) only by  a small amount.
 
For cycle pairs with short periods ($L < 50$), the non-zero distances
approximately follow a gaussian distribution with mean $\bar{d} =
0.45$ and standard deviation $\sigma_d\sim 0.15$ (see
Figure~\ref{fig:DiffHisto}, left column). Cycles with larger periods
display a non-gaussian distribution (see Figure~\ref{fig:DiffHisto},
center column) with positive skew, mean $\bar{d} \sim 0.18$, and
standard deviation $\sigma_d \sim 0.12$.  

Splitting the range of compared periods into smaller intervals
reveals that within these smaller intervals the distribution is
gaussian as well, but with decreasing mean for increasing periods (not shown).
Long-period limit cycles often tend to have a large fraction of their
neurons firing very close to 50\% of the time, reducing the average
distance.

\section{Performance of the random asymmetric neural network}
\begin{figure}[t]
  \centering
  \begin{minipage}[b]{0.48\textwidth}
    \centering
    \includegraphics[width=\linewidth]{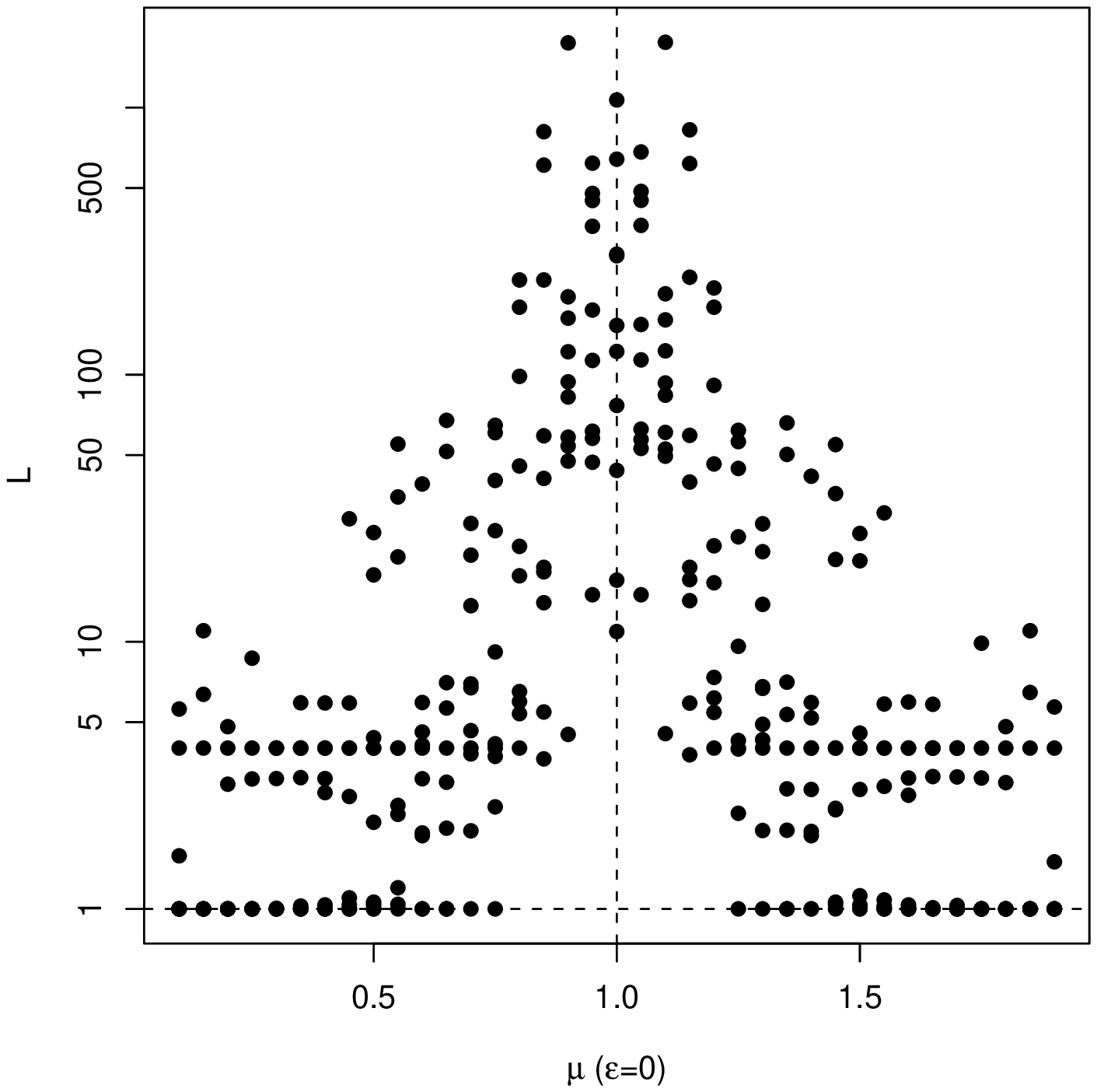}
    \caption[Cycle Length vs. Mean Bias]{The period $L$ of observed
      limit cycles as a function of the mean threshold level $\mu$,
      for $\eps=0$.  ($N = 40$ neurons; also note the semi-logarithmic
      L-scale).}
    \label{fig:Period_Mean}
  \end{minipage}%
  \hfill
  \begin{minipage}[b]{0.48\textwidth}
    \centering
    \includegraphics[width=\linewidth]{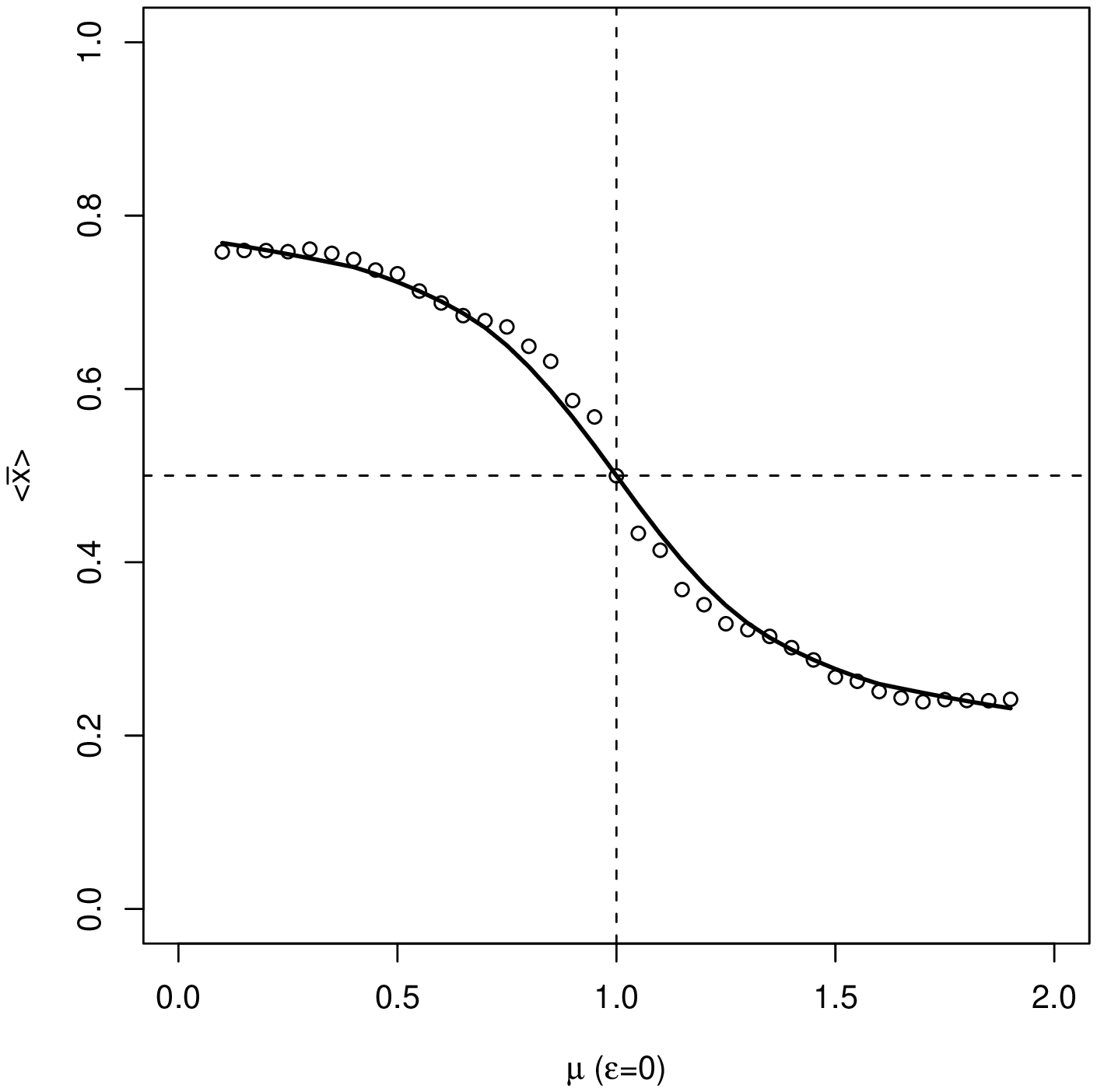}
    \caption[Mean Activity vs. Mean Bias]{The mean neural firing rate
     $\bar{x}$ (averaged over time and neuron index)
      for each observed limit cycle
      as a function of $\mu$, for $\eps=0$ and $N=40$.}
    \label{fig:FireFrac_Mean}
  \end{minipage}%
\end{figure}

The firing rate of random asymmetric neural networks tends to
converge to a fixed point
where the neural firing vector $\vec{x}(t)$ does not change in time,
if the deviation from normal thresholds is large. If the mean
threshold value is much greater or much less than normal, the neurons are
less or more likely to fire and the RSANN will tend to have a fixed point
with very few or very many neurons firing each time step.  These
extreme conditions are called network `death' and `epilepsy', respectively.
 By contrast, for normal thresholds $(\mu=1)$,
limit cycles with very long periods are possible, as seen in
Figs.~\ref{fig:Period_Mean}, \ref{fig:FireFrac_Mean} (cf.,
Clark, K\"{u}rten \& Rafelski (1988), K\"{u}rten (1988),
Clark (1990,1991), McGuire, Littlewort \& Rafelski (1991), McGuire {\it et al.} 
(1992)).

\begin{figure}[!h]
  \centering
  \begin{minipage}[b]{0.75\textwidth}
    \centering
    \includegraphics[width=\linewidth]{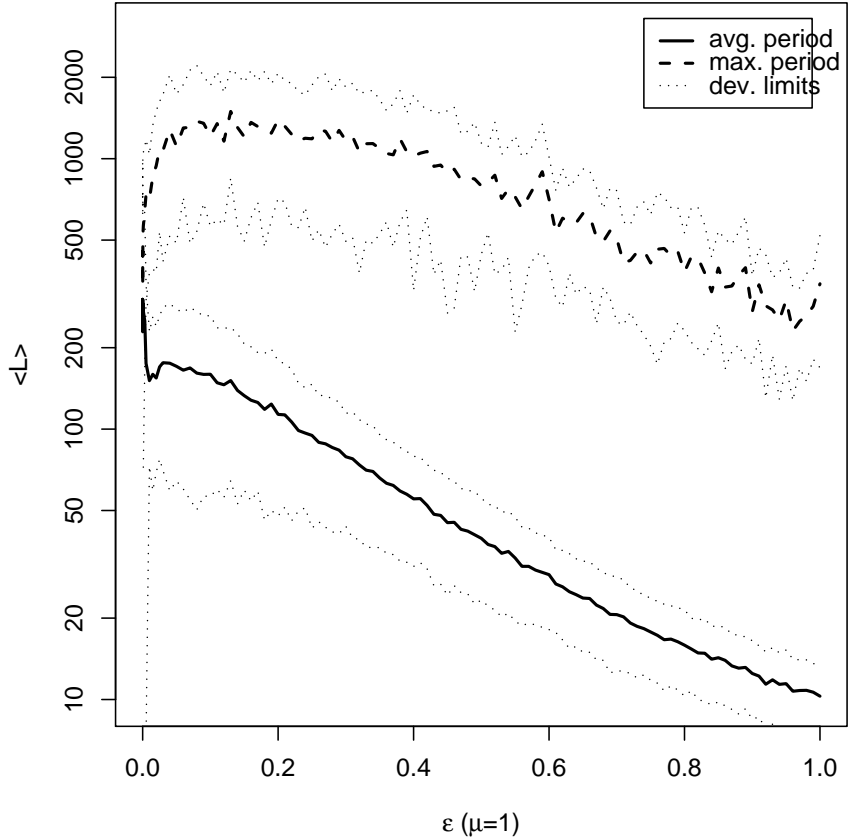}
    \caption[Mean Cycle Length vs. Noise Amplitude]{The period of
      the observed limit cycles, averaged over 100 trials, 
      decreases roughly exponentially as the noise amplitude $\eps$
      increases (for $\mu = 1$, $N = 50$; note the semi-logarithmic scale.)
      The average period for 10 networks is plotted as a solid line,
      and the average of the maximum periods for these 10 networks is plotted
      as a thick dashed line, with $1 \sigma$ deviation limits plotted
      as dotted lines.}
    \label{fig:Period_Eps}
  \end{minipage}%
\end{figure}

When the mean threshold value is normal ($\mu = 1)$, but the
threshold fluctuations from normality are large ($\eps >
\eps_2$), then there also exist {\em many} different mixed
death/epilepsy fixed points in which a fraction of the neurons are
firing at each time step and the remaining neurons never fire. With
growing $\eps$, this fraction of neurons with constant firing state ($x_i(t)=0$
or $x_i(t)=1$ for all $t$) grows, due to the fact that a larger fraction of
the neuron thresholds differ significantly from their normal values.
Therefore, as $\eps$ increases, a smaller fraction
of neurons actively participate in the dynamics,
making the effective network size smaller and the limit cycles shorter.
As can be seen in Figure~\ref{fig:Period_Eps}, the average period
and the maximal period both decrease roughly exponentially
with growing $\eps$. 

\begin{figure}[!h]
  \centering
  \hfill
  \begin{minipage}[b]{\textwidth}
    \centering
    \includegraphics[width=0.65\linewidth]{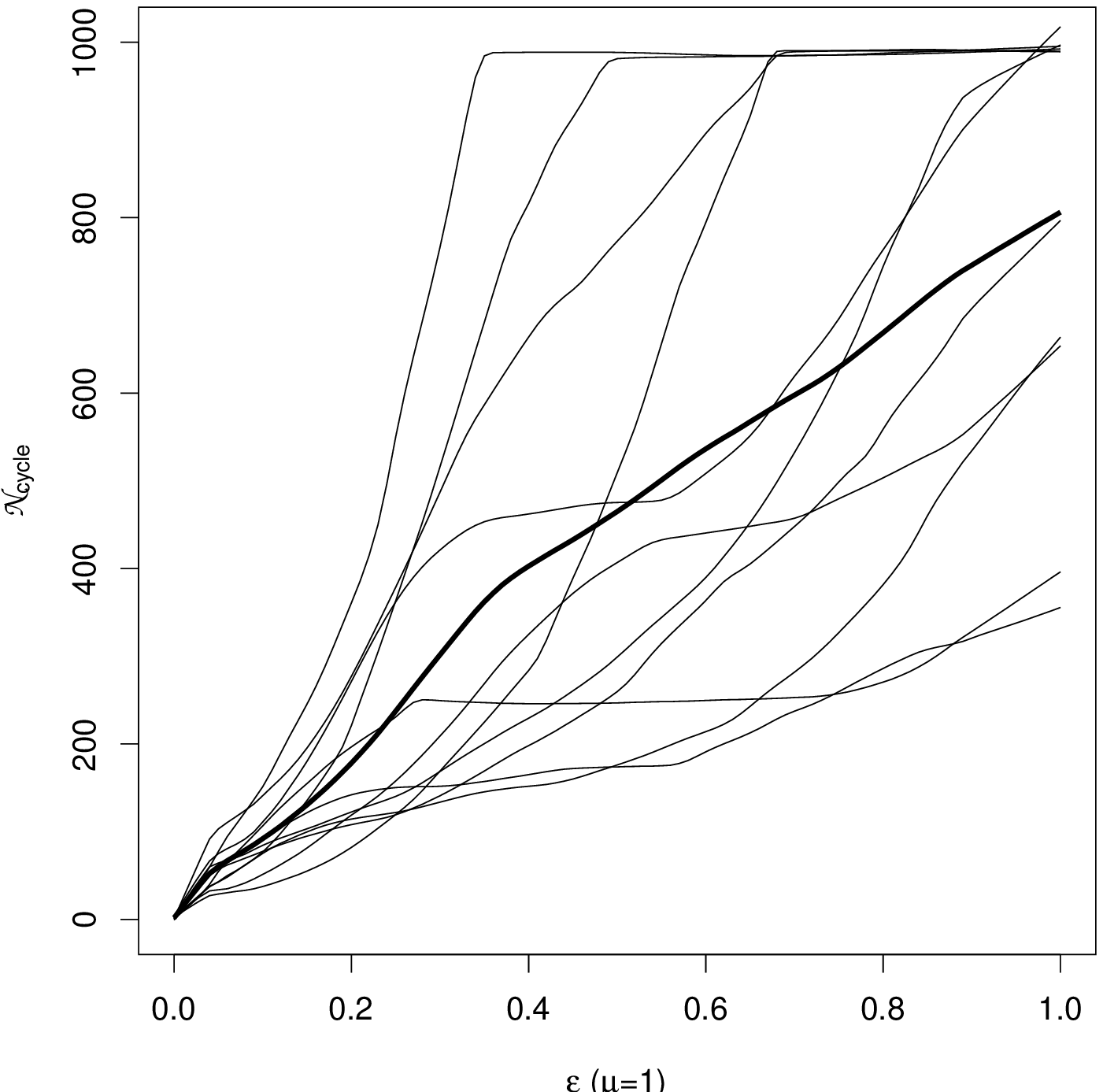}
    \includegraphics[width=0.65\linewidth]{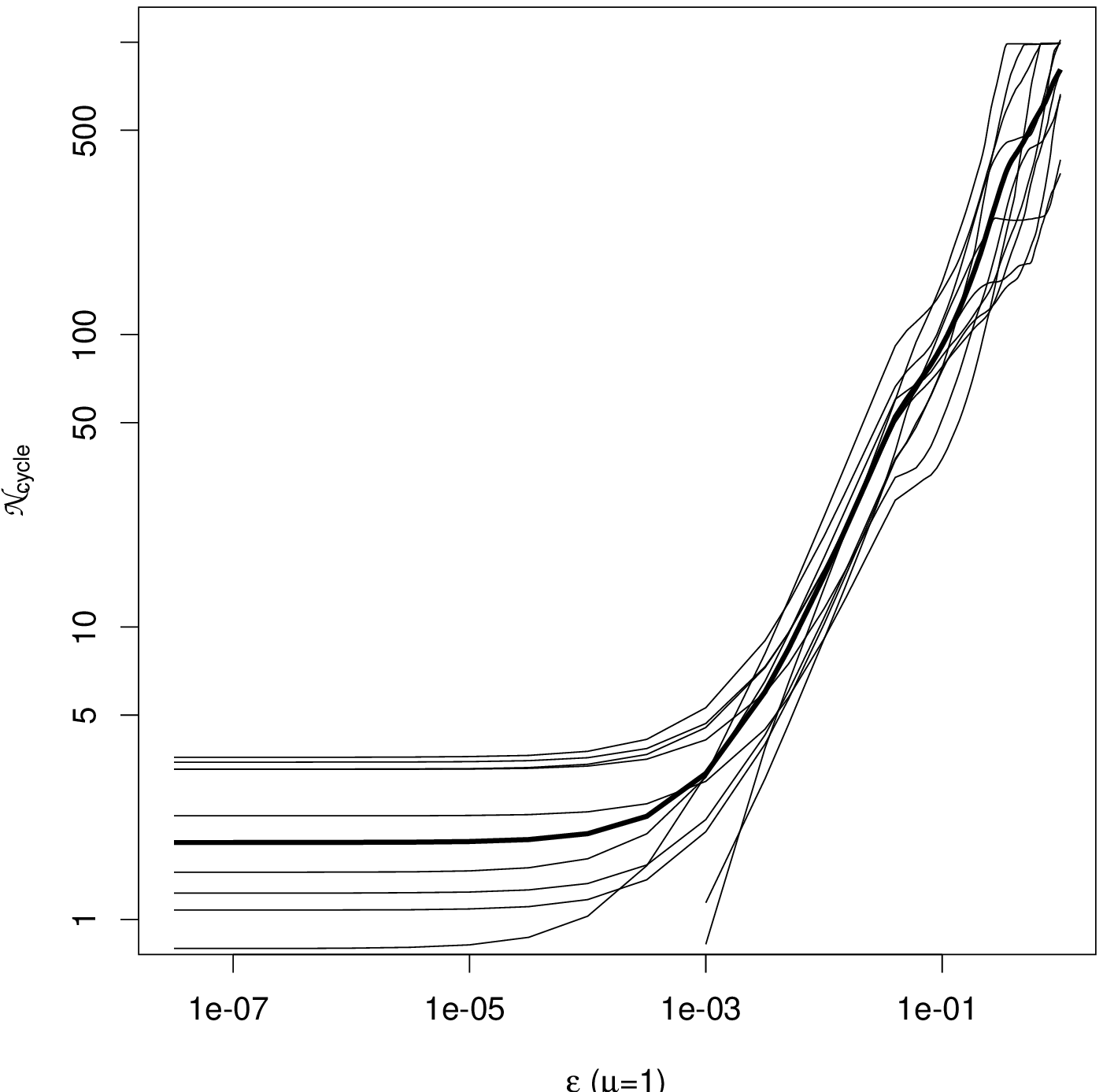}
    \caption[Number of Different Cycles vs. Noise Amplitude]{The
      total number of different limit cycles observed increases
      as the noise amplitude $\eps$ increases (for $N=50$, and
      for a maximal number of trials  $= 1000$,
      in linear and log-log plots).
      For $\eps>10^{-3}$, the
      ensemble-average scaling is approximately as a power law,
      ${\cal N}_{\text{cycles}} \sim \eps^{\alpha}$, with $\alpha \sim 1.0$,
      making it roughly a linear function as well.}
    \label{fig:NewCycles}
  \end{minipage}%
\end{figure}

Conversely, as $\eps$ increases, the number of different limit cycles
increases as well (as observed during many different trials, each with
a different random realization of thresholds $V_i \equiv \eta_i
V_i^0$).  This increase in the observed number of different cycles is
caused by each network realization eventually producing a (short)
limit cycle in a different portion of the network, as more and more
different neurons drop out of the picture as dynamical participants
(Figure~\ref{fig:NewCycles}). The saturation for some of the nets in
the ensemble of 10 nets is artificial, since we limited the maximum
number of trials to 1000 to constrain computational costs. 
\footnote{
In the accompanying log-log version of Figure~\ref{fig:NewCycles},
the behavior seems much more regular, with the large deviations from
the ensemble-average behavior for large $\eps$ becoming less important.
The general shape for $\eps > 10^{-3}$ is of a power law with exponent
near $1.0$ and positive coefficient (${\cal N}_{\text{cycles}}
\sim A \eps^{\alpha}$, with $\alpha \sim 1.0$), the near-unity exponent
of the power law making it roughly a linear dependence as well. For particular
examples from this 10-network ensemble, the dependence of the
${\cal N}_{\text{cycles}}$ curve is sometimes not a power-law;
and for the cases in which the behavior is similar to that of a power law,
the coefficient $A$ and exponent $\alpha$ of the power law both differ
by as much as a factor of 2 from the coefficient and exponent for the
average behavior.
The fact that the average behavior is a power law or even a linear
function (rather than irregular behavior) means 
that it might be worthwhile and interesting in the future
to perform a theoretical analysis of cycle diversity for RSANNs
as a function of threshold disorder.
}

In the following three sections, we limit our discussion to a small
ensemble of 10 networks with different connection-strength matrices,
and we compute the mean and variation of the quantities-of-interest,
as a function of the disorder amplitude, $\eps$.  In
section~\ref{sec:Ensemble}, we explore a much larger ensemble of 300
networks, but for only a few values of $\eps$.


\subsection{Eligibility}
\label{sec:Elig}
Since we are interested in complex dynamical behaviour, a large
fraction of neurons should participate non-trivially
in the dynamical collective
activity of the network -- such a network is said to have a high
degree of eligibility. A limit cycle $\alpha$ will have a
maximally eligible time-averaged firing pattern
$\bar{\vec{x}}^{\alpha}$, if $\bar{x}_i^{\alpha}=\frac{1}{2}$ for all
neurons $i$. There will be minimal eligibility if $\bar{x}_i^{\alpha} \in
\{0, 1\}$ for all neurons $i$. The Shannon information (or entropy)
has these properties, so we will adopt an entropy function
as our measure of the eligibility of a given limit-cycle attractor
$\alpha$:
\begin{equation}
  e(\alpha) \equiv 
  -\frac{1}{N} \sum_{i=1}^N \bar{x}_i^{\alpha} \ln \bar{x}_i^{\alpha} . 
\end{equation}
The mean eligibility ${\cal E}$, averaged over all 
${\cal N}_{\text{trials}}$ trials, is
\begin{equation}
  {\cal E} \equiv \frac{1}{{\cal N}_{\text{trials}}}
  \sum_{\alpha=1}^{{\cal N}_{\text{trials}}}
  e(\alpha) \leq \frac{1}{2} \ln 2 \equiv {\cal E}_{\text{max}} 
\end{equation}
for fixed network connectivity and fixed $\eps$.
Despite its utility, we do not have a rigorous dynamical motivation for
quantifying eligibility by entropy. As discussed at 
the beginning of this section, the fraction of actively participating
neurons decreases with growing $\eps$; thus eligibility is decreasing
as well (Figure~\ref{fig:Elig_Eps}). In other words, when the thresholds become
grossly `out-of-tune' with the mean membrane potential, the RSANN attractors
become more trivial, with each neuron tending toward its own independent
fixed point $x_i(t)=1$ or $x_i(t)=0$.

\subsection{Diversity}
\label{sec:Div}
We measure the accessibility of a given attractor by estimating the
probability $P(\alpha)$ that a given attractor (first observed at
trial $\alpha$) is observed during all ${\cal N}_{\text{trials}}$
trials, identified with its relative frequency of occurence 
\begin{equation}
  P(\alpha) \equiv \frac{{\cal N}(\alpha)}{{\cal N}_{\text{trials}}} \; ,
\end{equation}
where ${\cal N}(\alpha)$ is the number of observations of limit cycle
$\alpha$. Note that if a given
attractor is only observed {\em once}, then $P(\alpha) = 1/{\cal
  N}_{\text{trials}}$, while if the same attractor is observed in every 
trial, then $P(\alpha)=1$.

\begin{figure}[hp]
  \centering
  \subfigure{\includegraphics[width=0.38\linewidth]{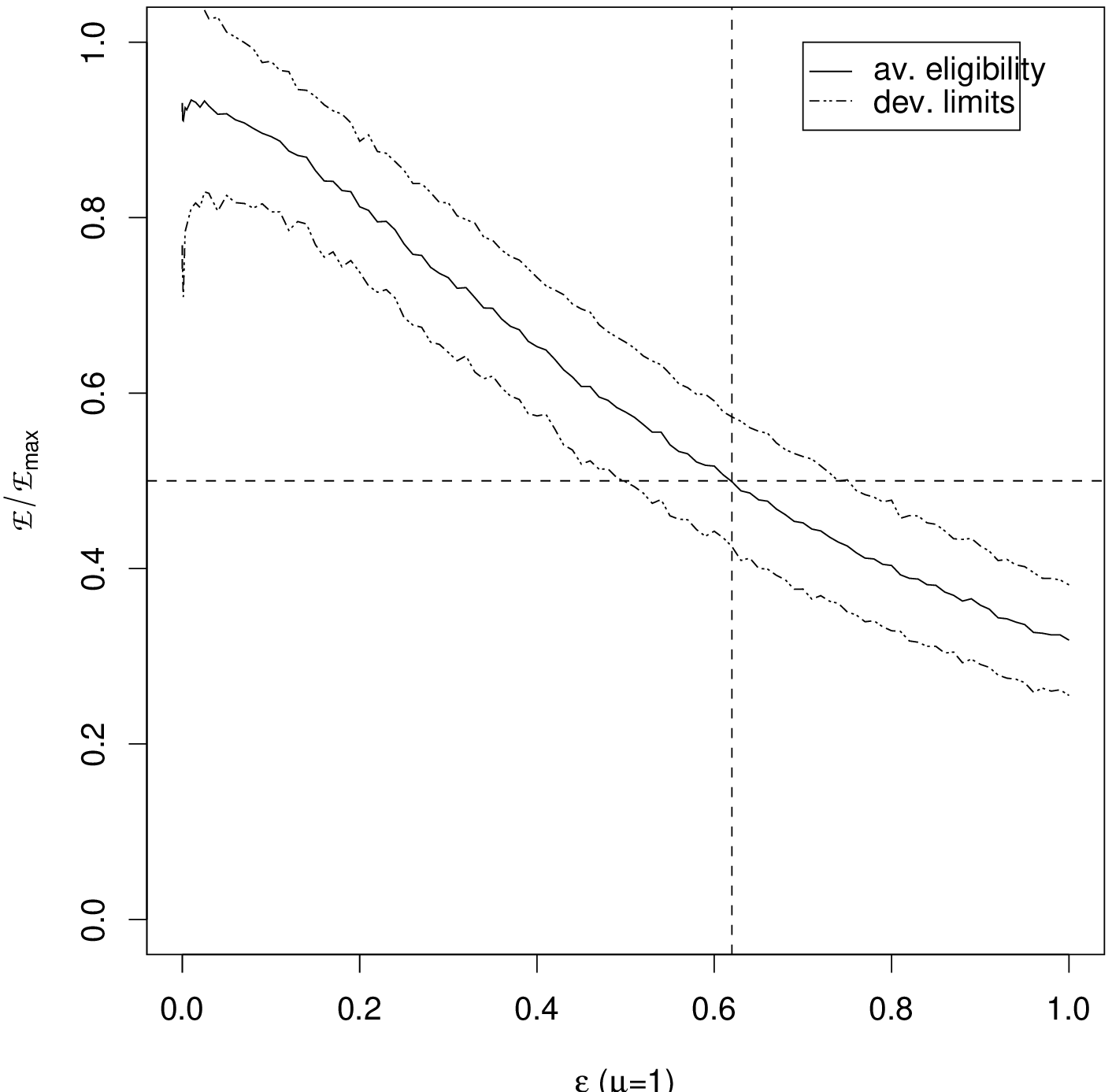}}
  \subfigure{\includegraphics[width=0.38\linewidth]{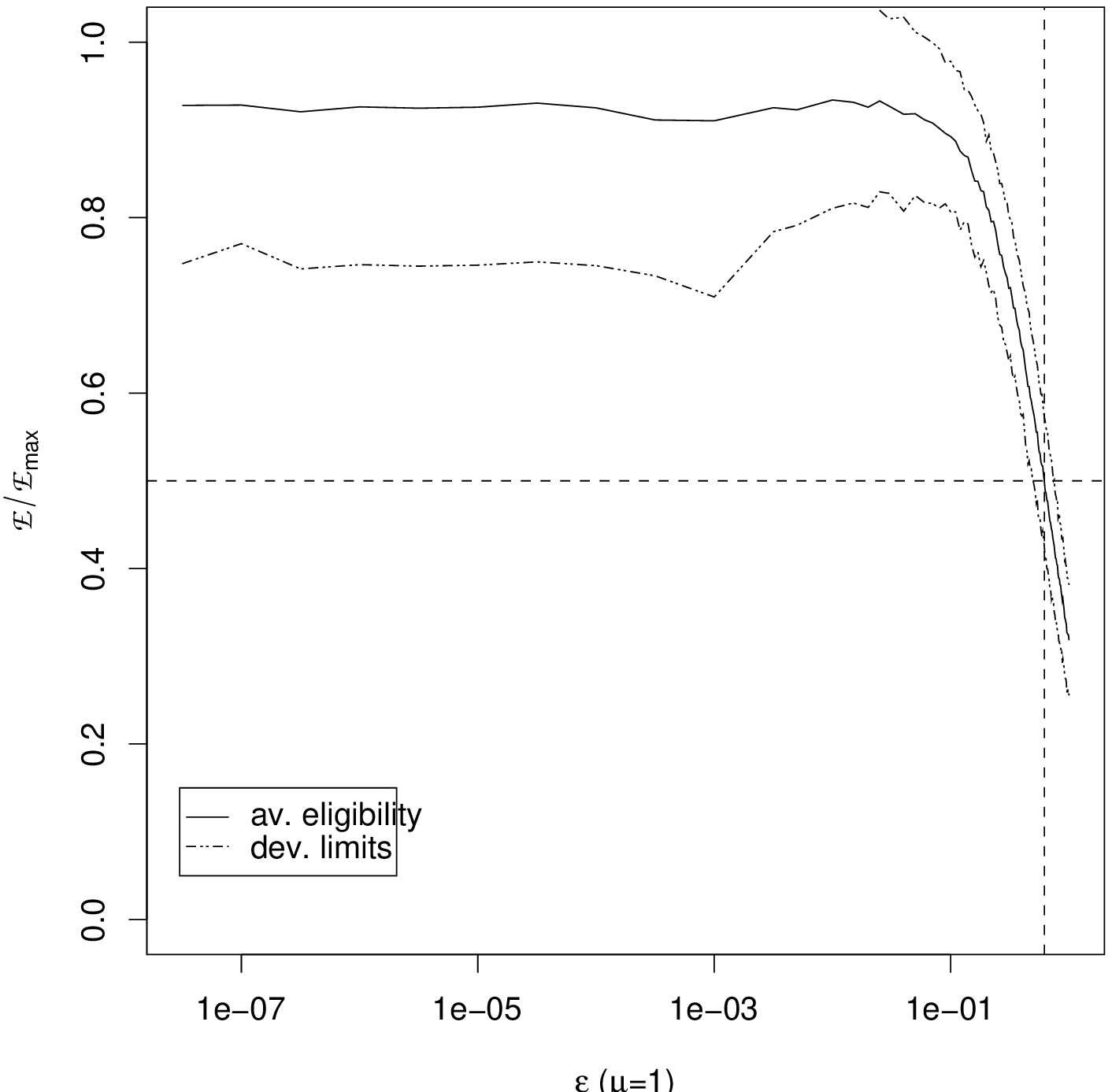}}
  \caption[Normalized Eligibility vs. Noise Amplitude]{
The mean eligibility ${\cal E}$ decreases with growing
noise amplitude $\eps$.  We present two views for Figs. 8-10,
 one with a semi-logarithmic scale, so that both the small-$\eps$
and large-$\eps$ behavior can be inspected 
($N=50$ for Figs. 8-10). }
  \label{fig:Elig_Eps}
  \subfigure{\includegraphics[width=0.38\linewidth]{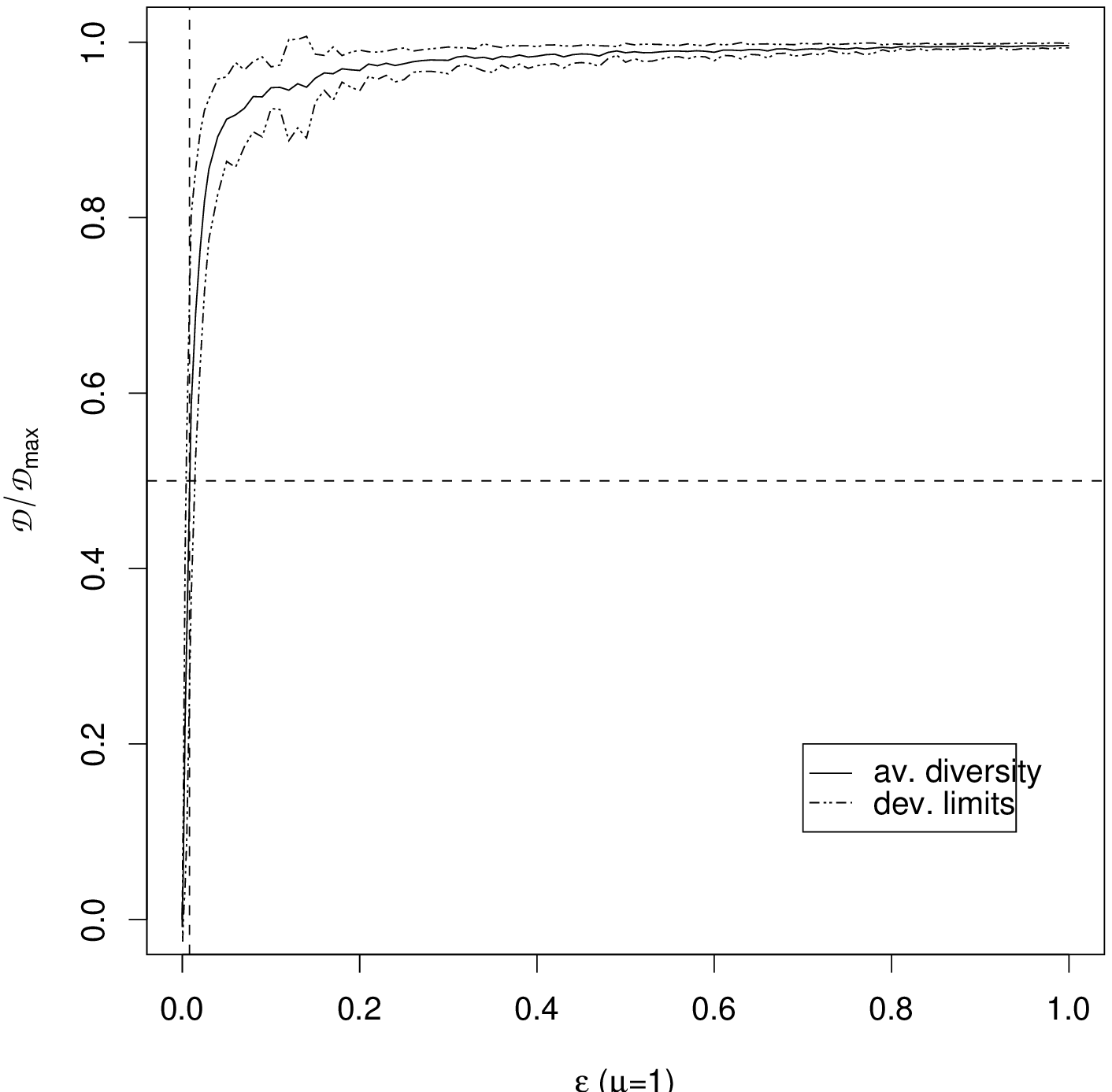}}
  \subfigure{\includegraphics[width=0.38\linewidth]{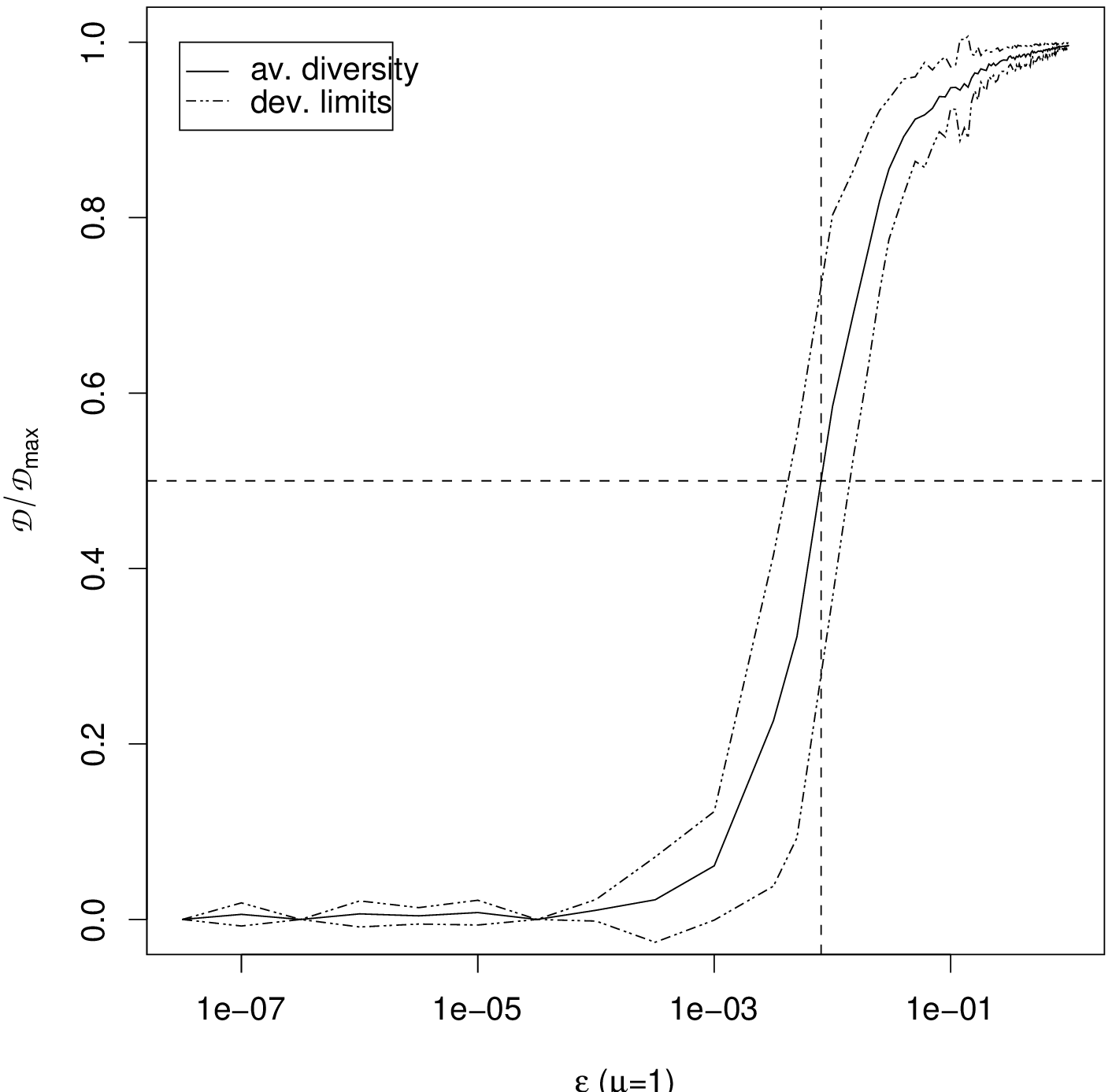}}
  \caption[Normalized Diversity vs. Noise Amplitude]{The
    diversity increases rapidly with growing noise amplitude $\eps$
(note the log scale for $\eps$ in one of these views).}
  \label{fig:Div_Eps}
  \subfigure{\includegraphics[width=0.38\linewidth]{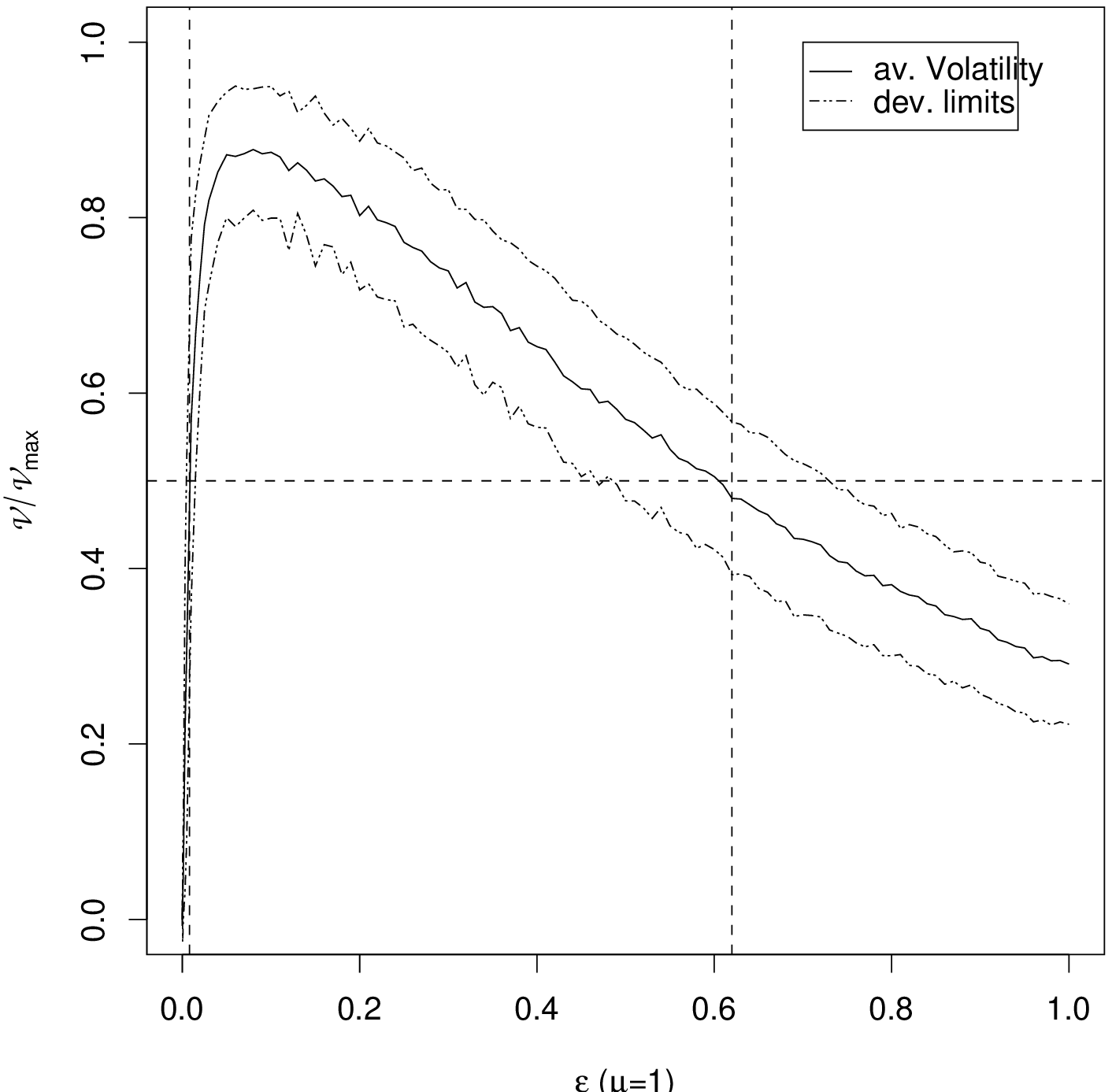}}
  \subfigure{\includegraphics[width=0.38\linewidth]{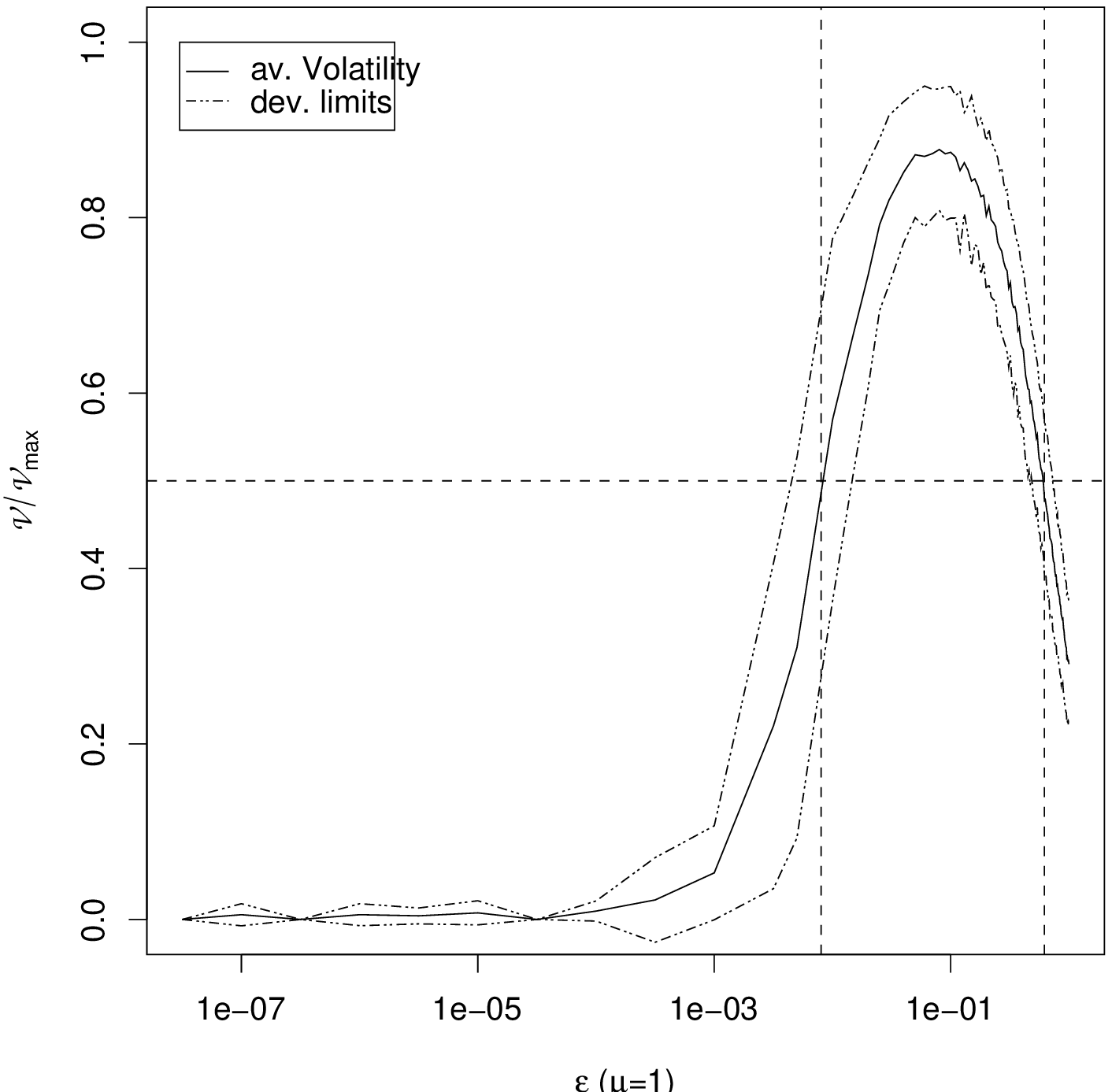}}
  \caption[Normalized Volatility vs. Noise Amplitude]{The volatility
    increases rapidly with growing noise amplitude due to the
    increasing diversity, as seen with both a linear and logarithmic
    scale in $\eps$.  It is large and nearly maximal within a rather
    broad range of noise amplitudes: $10^{-2} < \eps < 0.5$.}
  \label{fig:Vol_Eps}
\end{figure}

Given that each different limit-cycle attractor is accessed by the network
with probability $P(\alpha)$, we can define the diversity ${\cal D}$
as the attractor occupation entropy:
\begin{equation}
  {\cal D} = -\sum_{\alpha=1}^{{\cal N}_{\text{cycles}}} 
  P(\alpha)\ln P(\alpha) 
  \; \text{ , where} \quad 
  \sum_{\alpha=1}^{{\cal N}_{\text{cycles}}} P(\alpha) = 1 
  \label{eq:Div}
\end{equation}
and ${{\cal N}_{\text{cycles}}}$ is the total number of different
observed cycles.  It is easily seen that a large ${\cal D}$
corresponds to the ability to occupy many different cyclic modes with
nearly equal probability; the diversity will reach a maximum value of ${\cal
  D}_{\text{max}} = \ln {{\cal N}_{\text{trials}}}$ when $P(\alpha) =
1 / {{\cal N}_{\text{trials}}}$ for all limit cycles $\alpha$, i.e.
if in each trial a different cycle is observed. A small value of ${\cal
  D}$ corresponds to a strong stability (or inflexibility)
 of the system -- very few
different cyclic modes are available. As can be seen from
Figure~\ref{fig:Div_Eps} the diversity grows rapidly with
increasing disorder amplitude $\eps$, and with a relatively small
disorder value of $\eps = \eps_1 \sim 10^{-2}$,
the diversity is already half of its maximal value. 
Since it takes into account the accessibility of all the detected cycles,
this technique of quantifying diversity by an entropy function is
considerably more robust and meaningful than simply counting the cycles.

\subsection{Volatility}
\label{sec:Vol}
Volatility is defined as the ability to access a large number of
highly eligible limit cycles, or a `mixture' of high eligibility and high
diversity. Having defined both eligibility and diversity, we can now
combine them to define volatility as an entropy-weighted entropy:
\begin{equation}
  {\cal V} = -\sum_{\alpha=1}^{{\cal N}_{\text{cycles}}} 
  e(\alpha) P(\alpha) \ln P(\alpha) \; .
\end{equation}
Since the volatility curve in Figure~\ref{fig:Vol_Eps} is roughly
the product of the eligibility and the diversity curve in
Figures~\ref{fig:Elig_Eps} and \ref{fig:Div_Eps}, there exists an
intermediate regime $\eps_1 < \eps < \eps_2$ of high volatility. At
$\eps = \eps_1 \sim 10^{-2}$ the growing disorder amplitude causes a
transformation to a condition of diversity,
entailing many different limit cycles, whereas at
$\eps = \eps_2 \approx 0.5$ the disorder amplitude has become
so large that most limit cycles become fixed points. We accordingly
identify three different regimes for the RSANN with disorder:
\begin{enumerate}
\item Stable Regime: $\eps < 10^{-2}$
\item Volatile Regime: $10^{-2} \leq \eps \leq 0.5$
\item Trivially Random Regime: $\eps > 0.5$.
\end{enumerate}

\subsection{Larger Ensemble study}
\label{sec:Ensemble}

\begin{figure}[hp]
  \includegraphics[width=0.49\textwidth]{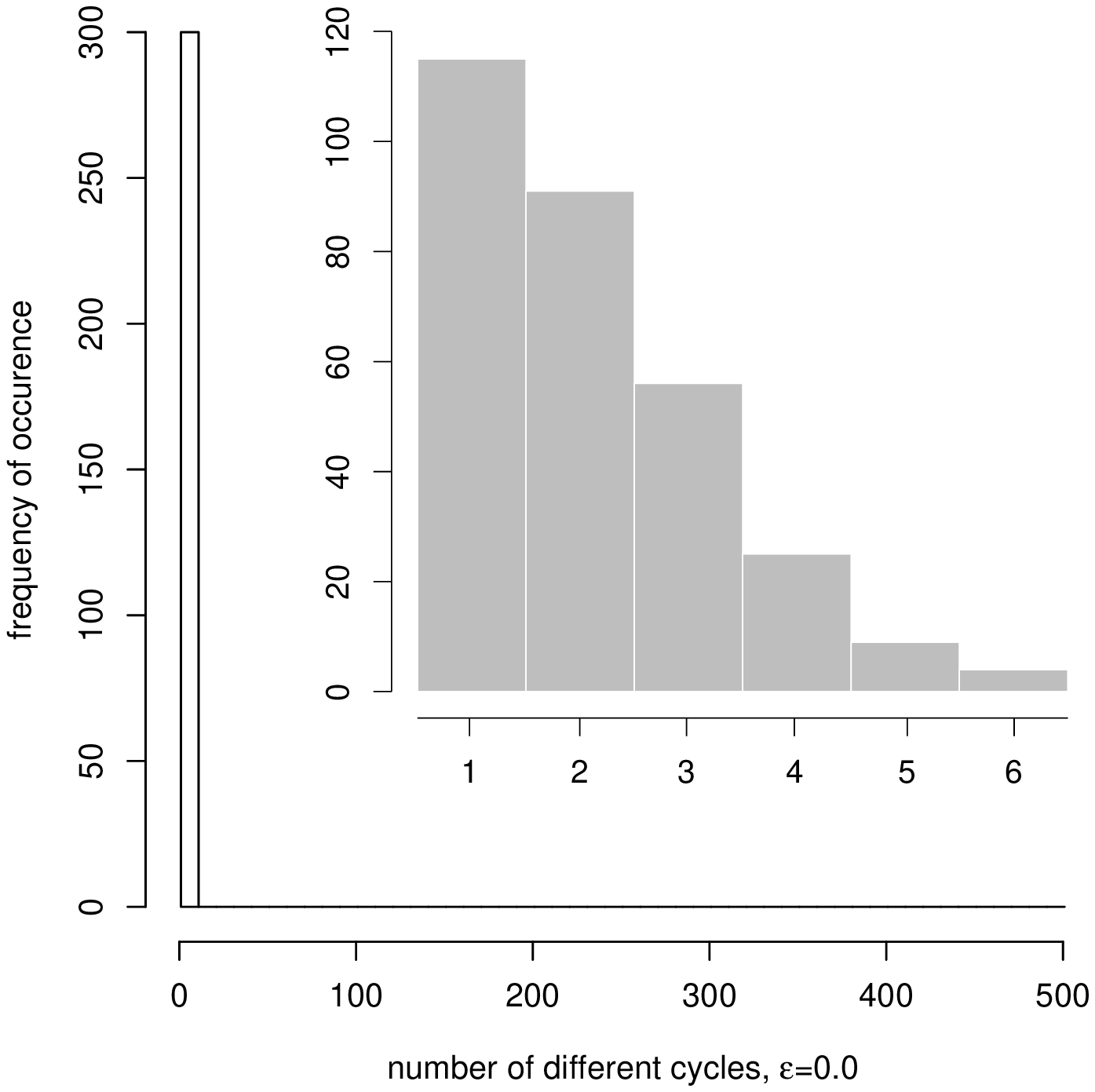}
  \includegraphics[width=0.5\textwidth]{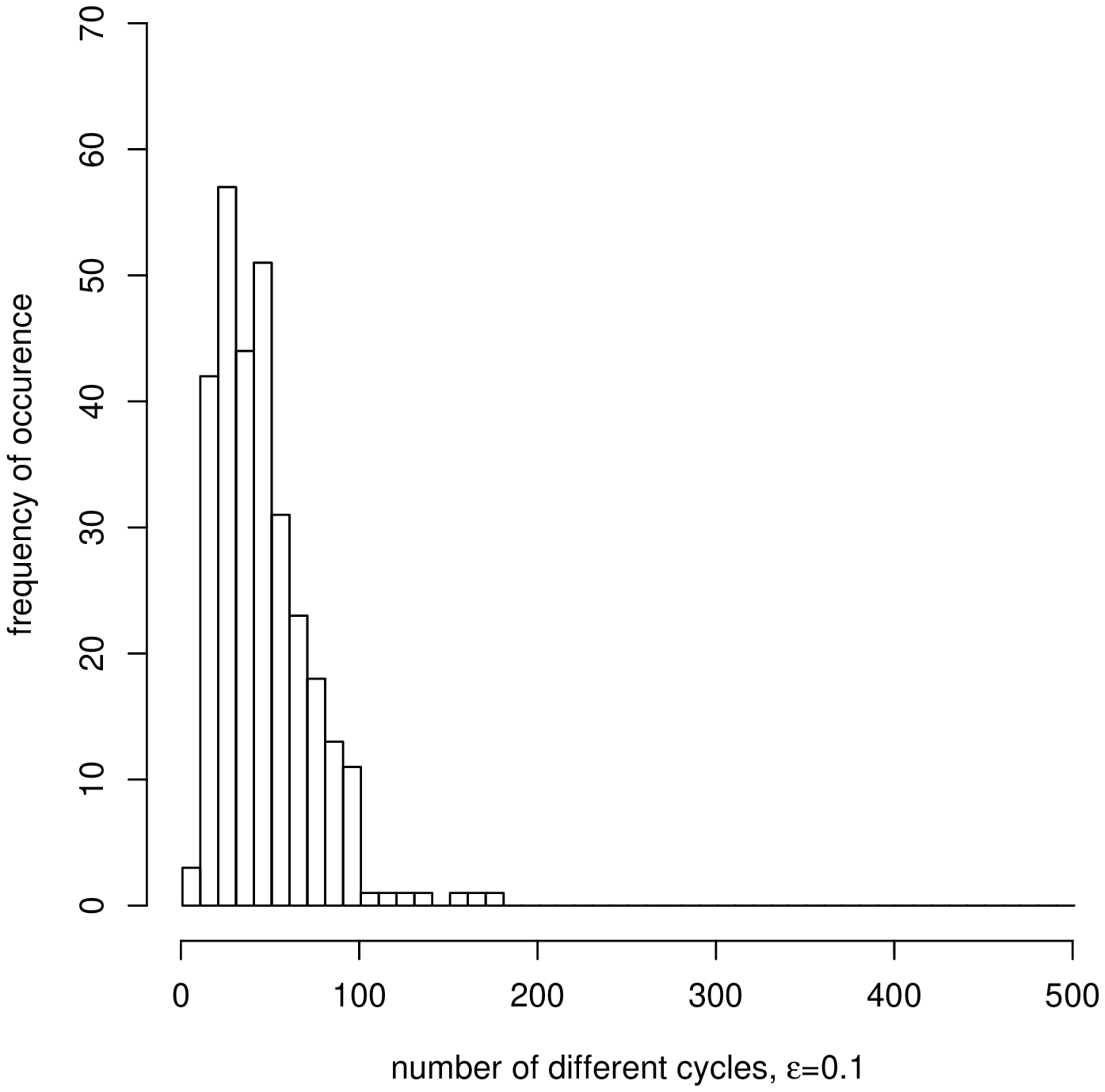}
  \includegraphics[width=0.49\textwidth]{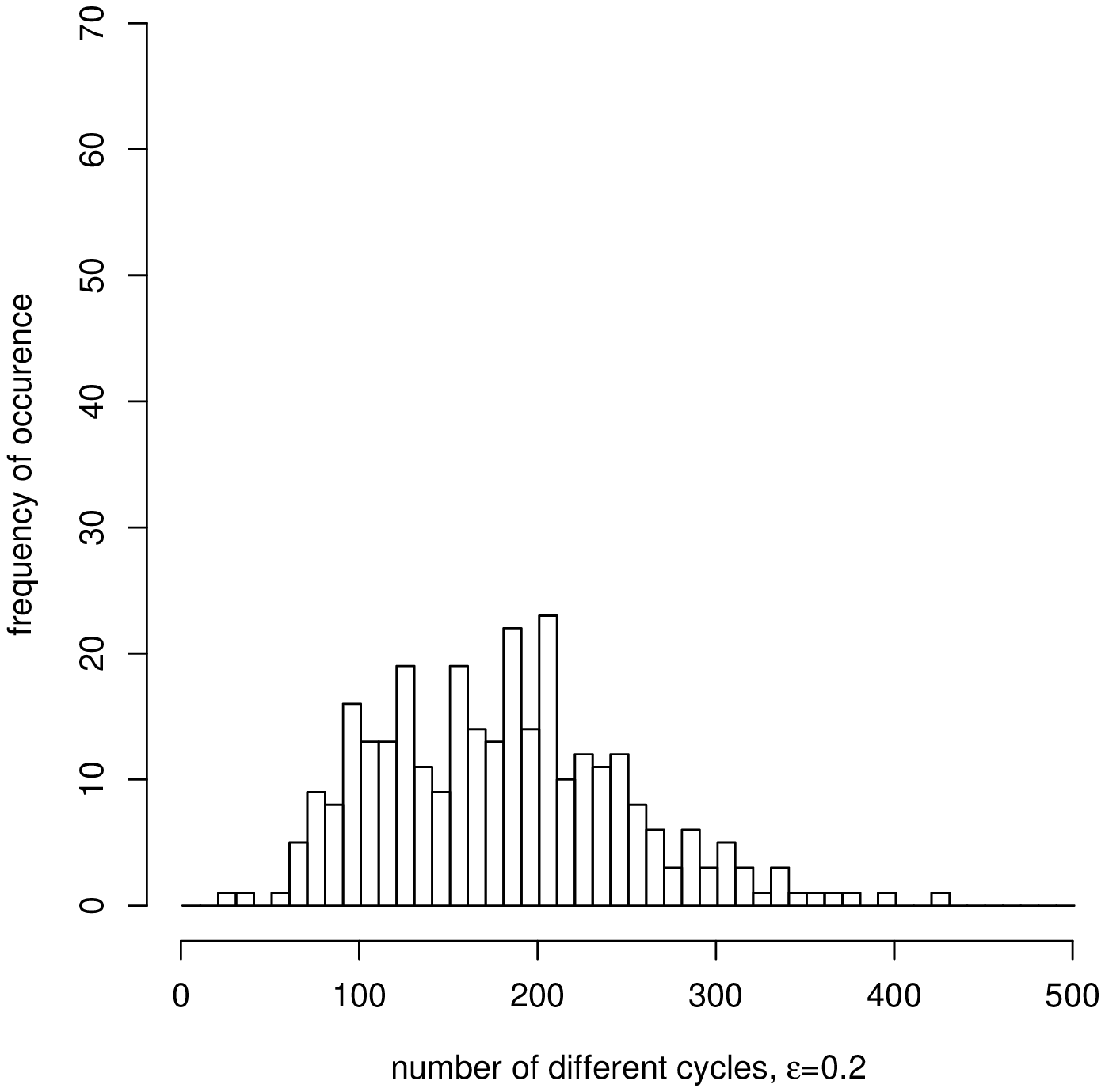}
  \includegraphics[width=0.5\textwidth]{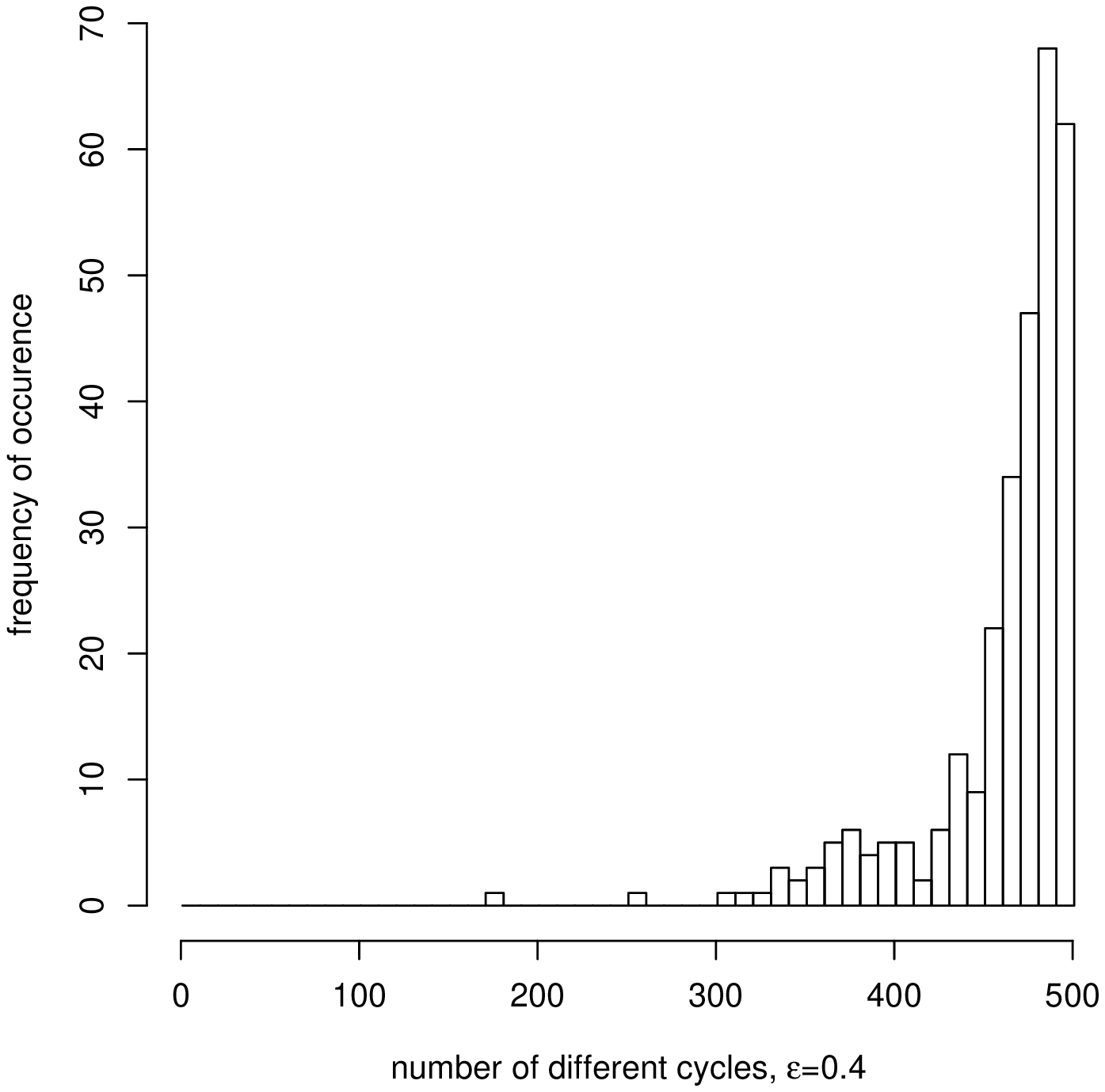}
  \caption{
For 300 networks of $N=50$ neurons, each with different connection strength
matrices, we form histograms of
the number of limit cycles found in each different network.
Histograms for four
different values of $\eps$ are presented. Note the rebinned histogram
in the inset of the $\eps = 0.0$ histogram.
}
\label{fig:histonumdiff}

\end{figure}

The results obtained in Sections~\ref{sec:Elig}~-~\ref{sec:Vol}
were obtained from ten RSANNs with random weight matrices drawn from
the uniform distribution $w_{ij} \in [-1,1]$. As can be seen from the
standard deviation curves in figures~\ref{fig:Elig_Eps}-~\ref{fig:Vol_Eps},
which are in close proximity to the average curves, all networks
exhibit the same qualitative behaviour.
To further confirm this finding, we have tabulated in
tables~\ref{tab:numfig} \&~\ref{tab:periodlength} and
Figure~\ref{fig:histonumdiff}
the statistics of the number of different
limit cycles, their periods, and the diversity \& volatility,
for 300 networks with different connection strength matrices
using 500 different disorder vectors $\vec{\eta}$ for each network.
The $\eps=0.0$ results in Table~\ref{tab:numfig} and
Figure~\ref{fig:histonumdiff} confirm Amari's theoretical result (1974)
and the empirical results of Clark, K\"{u}rten \& Rafelski (1988),
that random networks tend to possess only a small number of different cyclic
modes. For larger $\eps$, the ensemble-average number of cycles, 
the diversity and volatility are all much larger than for $\eps=0$.

In particular, the results shown in Table~\ref{tab:periodlength} suggest
that with increasing $\eps$, the minimum period rapidly decreases,
whereas the maximum period rapidly increases, with the average period
staying roughly constant and closer to the minimum than the
maximum. This suggests that at least a few 
limit cycles having long periods exist with non-normal thresholds, and
that there are many more fixed points than long-period limit cycles
when the thresholds are far from normal,
than when the thresholds are close to normal.

\begin{table}[th]
  \begin{tabular}{r|r|r|r|r|r}
    & \multicolumn{3}{|c|}{Number of different cycles (300 matrices)} \\

    $\eps$ & max. \# of cycles  & ave. \# of cycles & ave. \# of long cycles & diversity & volatility \\
    \hline
    0.0 &   6 &   2.11 $\pm$  1.17 &  0.04 $\pm$  0.19 &
    0.06 $\pm$ 0.08 & 0.03 $\pm$ 0.04 \\

    0.1 & 176 &  45.58 $\pm$ 26.54 &  1.59 $\pm$  2.12 &
    0.33 $\pm$ 0.14 & 0.26 $\pm$ 0.10 \\

    0.2 & 422 & 179.98 $\pm$ 69.87 & 12.66 $\pm$ 12.89 &
    0.70 $\pm$ 0.12 & 0.55 $\pm$ 0.12 \\

    0.4 & 500 & 461.34 $\pm$ 44.24 & 18.40 $\pm$ 19.89 &
    0.98 $\pm$ 0.04 & 0.64 $\pm$ 0.14 \\
  \end{tabular}
\caption{For 300 networks of $N=50$ neurons, each with a different
matrix of connection strengths, we tabulate the statistics of the number
of observed limit cycles (maximum number, average number of cycles,
and average number of long cycles),
as well as the average diversity and the average volatility. The
results are given for 4 different values of threshold disorder $\eps$.}
\label{tab:numfig}
\end{table}

\begin{table}[h]
\begin{tabular}{r|r|r|r}
  & \multicolumn{3}{|c}{Average period length (300 matrices)} \\
  $\eps$ & ave. min. period & ave. max. period & ave. mean period \\
  \hline
  0.0 & 64.98 &  111.97 &  85.63 \\
  0.1 &  2.10 &  796.63 &  89.18 \\
  0.2 &  1.27 & 1233.89 & 146.66 \\
  0.4 &  1.02 & 1144.64 & 110.15 \\
\end{tabular}
\caption{
For 300 networks of $N=50$ neurons, each a different matrix
of connection strengths,
we tabulate the averages of the statistics of the observed periods of the
limit cycles for each network (minimum period, maximum period, mean period),
 for four different values of the threshold disorder $\eps$.}
\label{tab:periodlength}
\end{table}

\subsection{Stability and attractor basins of observed cycles}
\label{sec:Stability}

We have shown in Sections~\ref{sec:Elig}~-~\ref{sec:Ensemble} that when
we change the threshold parameters of a network sufficiently far
from their default values, then we get new, non-trivial behavior for
nearly each parameter realization. Since
the level of threshold disorder which is needed
to obtain new and complex behavior is not too high, the ensemble of networks
can exhibit diverse and complex behavior with only slight changes
in the network parameters.

\begin{figure}[ht]
  \centering
  \begin{minipage}[b]{0.48\textwidth}
    \centering
    \includegraphics[width=\linewidth]{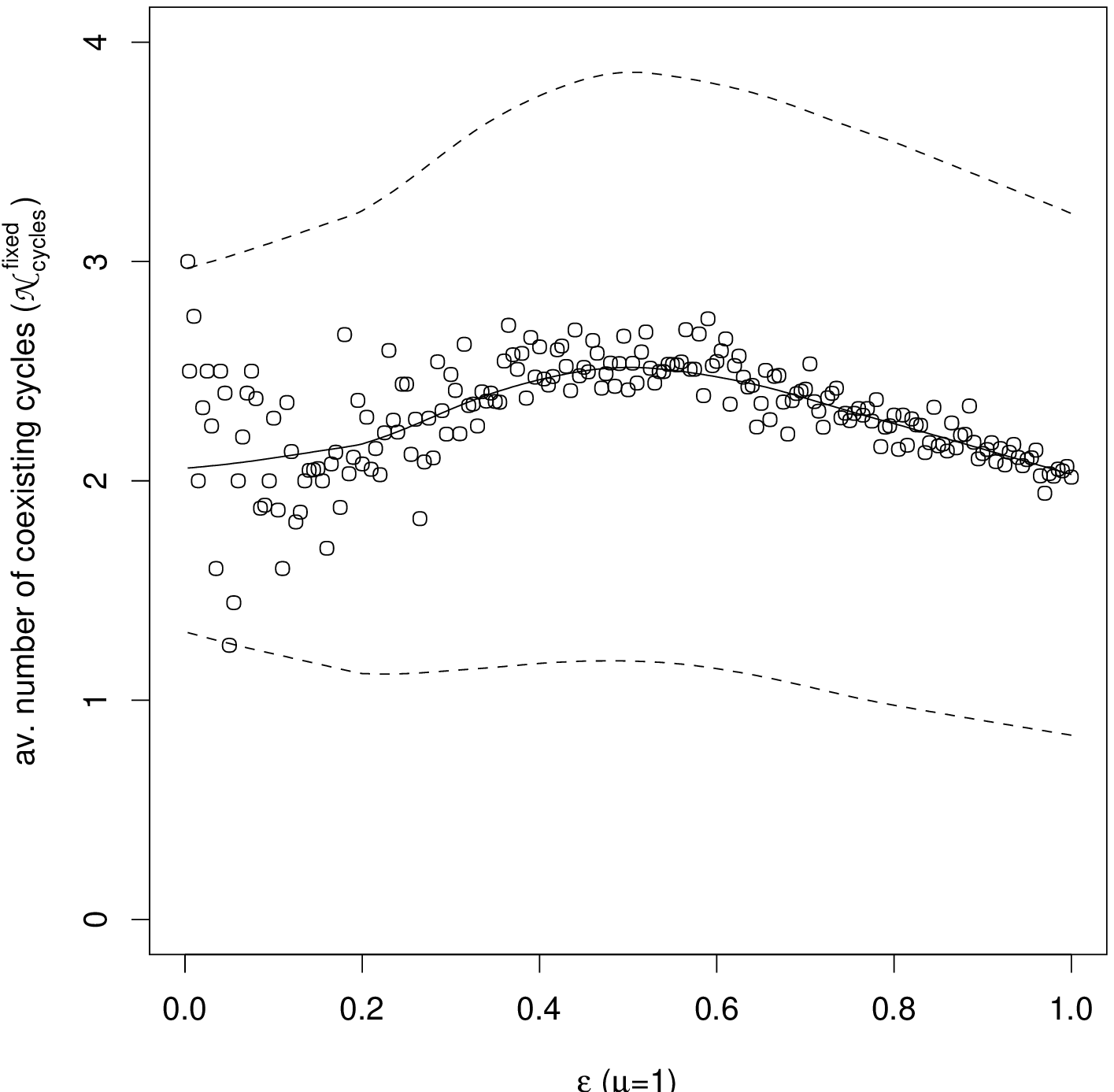}
    \caption[Number of Cycles per configuration vs. Disorder Amplitude]{%
The number of repeatedly observed limit cycles
with a fixed set of connection strengths and thresholds 
 but different initial conditions $\vec{x}(0)$ is relatively stable
with $N_{\text{cycles}}^{\text{fixed}} \sim 2\pm 1$.  
(for $N=50$ and $\mu=1$; with 100 random initial conditions,
averaged over several threshold realizations; the solid
curve is a smoothed average curve; the dotted curves
are smoothed $1\sigma$ error curves). }
    \label{fig:Stab_Cycles_Eps}
  \end{minipage}%
  \hfill
  \begin{minipage}[b]{0.48\textwidth}
    \centering
    \includegraphics[width=\linewidth]{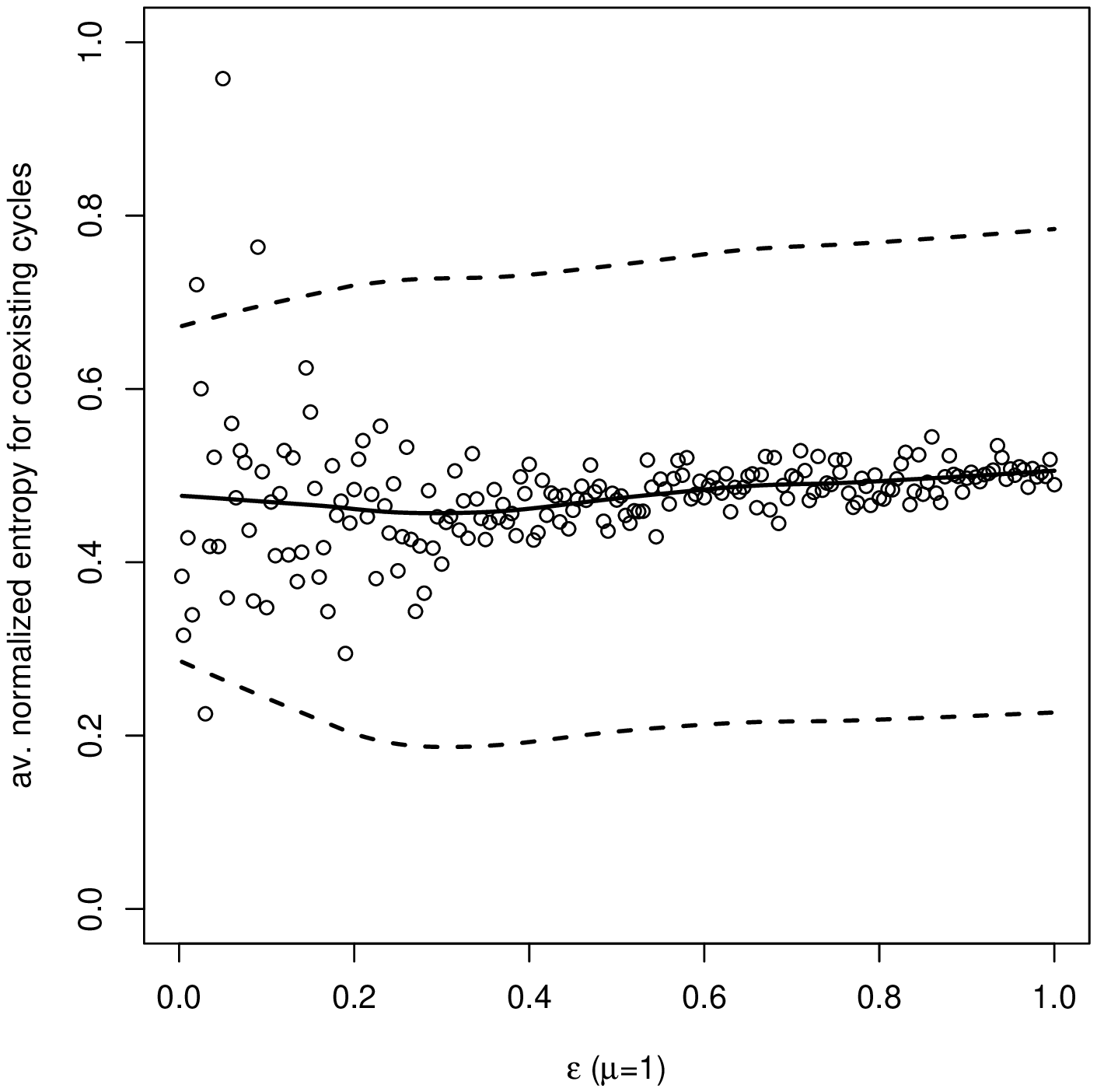}
    \caption[Number of Different Cycles vs. Disorder Amplitude]{%
      The attractor occupation entropy (Eq.~\ref{eq:Div}) of the small
      number of observed cycles (compare Fig.~\ref{fig:Stab_Cycles_Eps})
      allows us to estimate the relative sizes of the basins of attraction.
      Since the entropy reaches only half of its maximum, some
      cycles dominate over the others. The solid curve is a smoothed
      average curve; the dashed curves are smoothed $1\sigma$ error
      curves.}
    \label{fig:Stab_Entropy_Eps}
  \end{minipage}%
\end{figure}

In order to demonstrate that this diversity originates only
from ensemble diversity and
is not intrinsic to the specific network realizations, we must show
that each specific network of the ensemble possesses only a limited set of
limit cycles. For this reason we counted the number of different
observed cycles ${\cal N}_{\text{cycles}}^{\text{fixed}} (w_{ij},
\vec{\eta}(\eps))$ of each network $(w_{ij}, \vec{\eta}(\eps))$ in the
ensemble starting with 100 randomly-chosen initial activity patterns
$\vec{x}(0)$. Then we performed the ensemble average 
\begin{equation}
  {\cal N}_{\text{cycles}}^{\text{fixed}}(\eps) = 
  \langle
    {\cal N}_{\text{cycles}}^{\text{fixed}} (w_{ij}, \vec{\eta}(\eps))
  \rangle_{\vec{\eta}(\eps)}\,.
\end{equation}
As can be seen from Figure~\ref{fig:Stab_Cycles_Eps}, the number of
different observed cycles is rather small, {\it and} does not depend
upon the disorder amplitude $\eps$. Of course, the {\it actual} size
of the repertoire does depend on the instantiation of
$\vec{\eta}(\eps)$, as evidenced by the deviations (indicated by
dashed curves), but the degree of variation in ${\cal
  N}_{\text{cycles}}^{\text{fixed}}(\eps)$ in no way matches the
substantial secular increase of ${\cal
  N}_{\text{cycles}}^{\text{fixed}}(\eps)$ observed in
Figure~\ref{fig:NewCycles}.

Using a similar diversity measure like in Equation~\ref{eq:Div}, but
now using the occurence probabilities of observed cycles with fixed
network parameters, it is possible to estimate the relative sizes of
the basins of attraction of these cycles. The diversity becomes
maximal when all cycles are observed with equal probability,
corresponding to basins of attractions of equal size.  As can
be seen from Figure~\ref{fig:Stab_Entropy_Eps}, the diversity reaches
only the half of its maximum, indicating basins of attraction of different
sizes. The figure suggests that $-\langle\ln P\rangle \sim
\frac{1}{2}$, implying that $\langle P \rangle \sim 0.6$. In other words,
for fixed thresholds, as we vary the initial firing vector, we observe the
same cycle in greater than $\sim 60\%$ of the trials;
as evident in Figure~\ref{fig:Stab_Entropy_Eps}, there is
little dependence in the fixed-threshold cycle-diversity
upon the frozen-disorder amplitude $\eps$. 

\subsection{Distribution of Limit-Cycle Periods and their
Dependence upon Network Size}
\label{sec:NetSize}
\begin{figure}[ht]
  \centering
  \begin{minipage}[t]{0.48\textwidth}
  \centering
  \includegraphics[width=\linewidth]{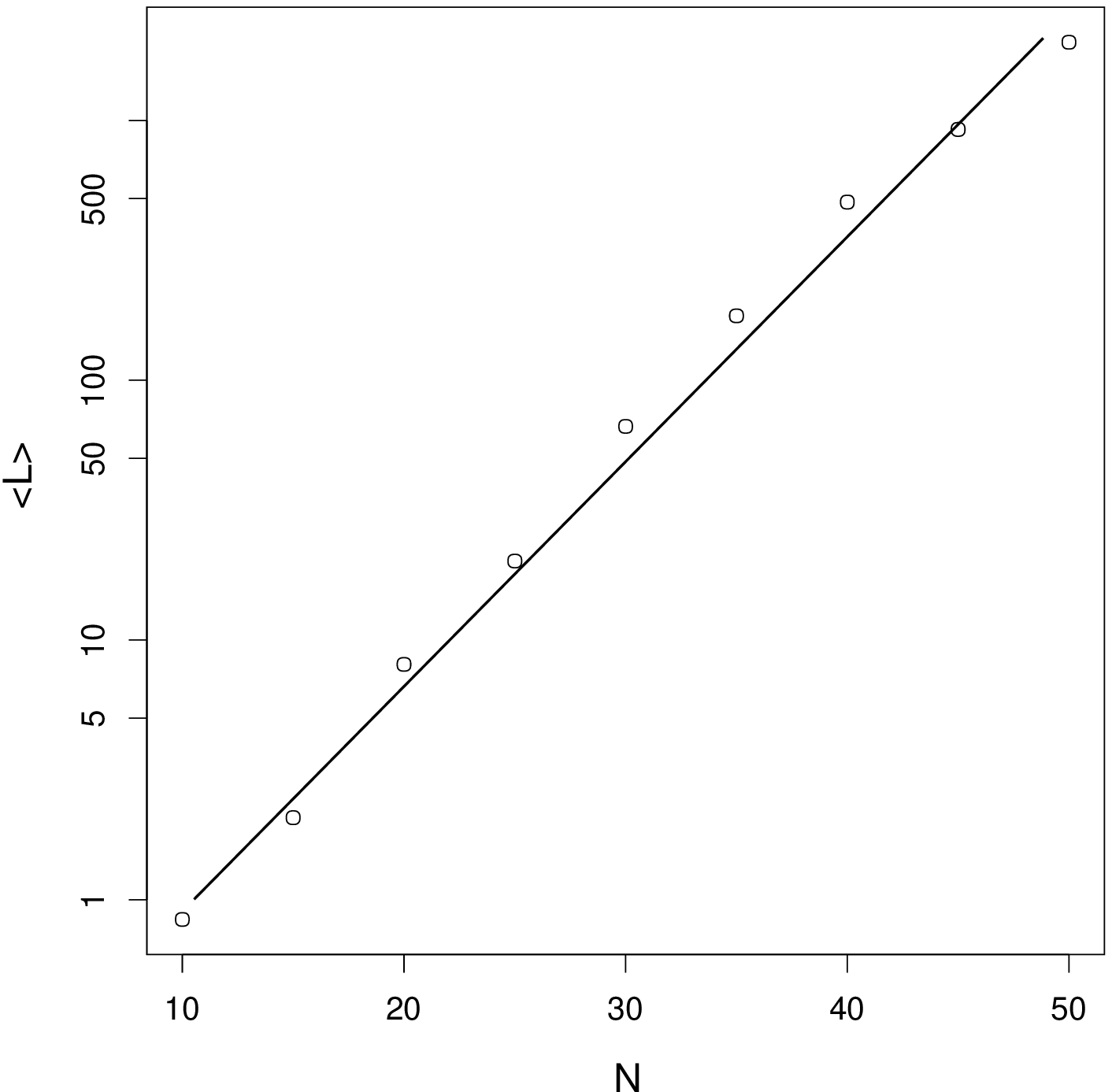}
  \caption{Mean cycle length $\langle L \rangle$,
averaged over 20 weight matrices with normal thresholds and $\eps=0.1$,
grows exponentially
with network size $N$. In Figure~\ref{fig:PeriodHisto},
we observe that the distribution
of limit-cycle periods for fixed $N$ has
a broad range of periods ranging from $L=1$ to roughly $2\langle L\rangle$.
Therefore, using standard error bars on the values of $\langle L \rangle$ in
the above plot would not be very informative.}
  \label{fig:PeriodLen_NetSize}
  \end{minipage}
  \hfill
  \begin{minipage}[t]{0.48\textwidth}
  \centering
  \includegraphics[width=\textwidth]{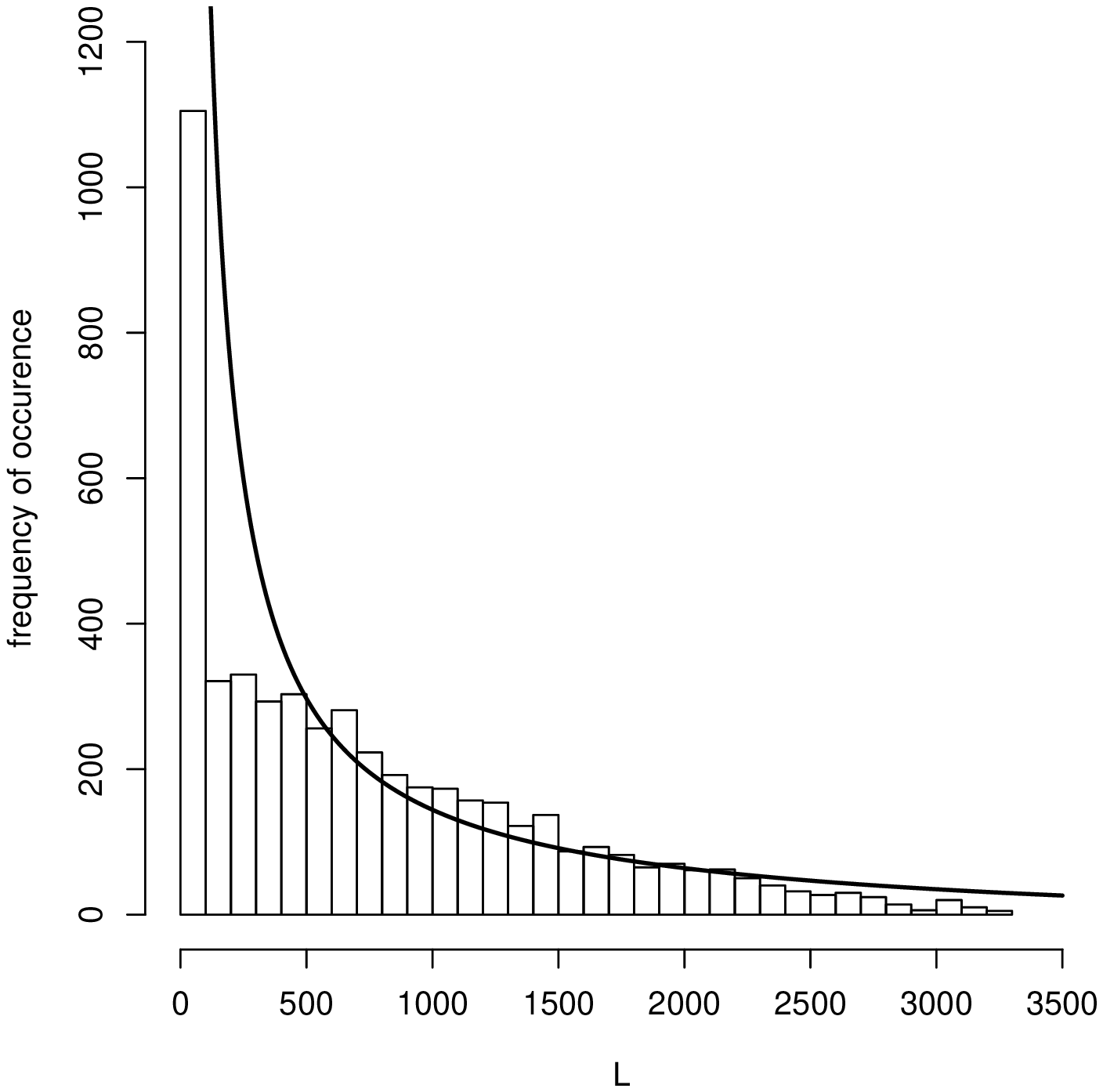}
  \caption{Distribution of different limit-cycle periods
 observed in the volatile regime ($N=50$, $\eps=0.02$).
 We overlay a fit to the distribution predicted by Bastolla \& Parisi
 (1997, 1997b) for Kauffman nets.}
  \label{fig:PeriodHisto}
  \end{minipage}
\end{figure}

 For the choice of threshold and connectivity parameters made here, the average
cycle length $\langle L \rangle$ grows exponentially with the network
size $N$ (see Fig.~\ref{fig:PeriodLen_NetSize}).  This
exponential scaling of the cycle length puts it in the `chaotic'
regime of K\"{u}rten's (1988) classification of dynamical phases, where the
motion shows high sensitivity to initial conditions. K\"{u}rten
also found a `frozen' regime where the limit-cycle
period scales as a power law in $N$, and where there is
little sensitivity of the attracting limit cycle upon initial conditions.  When
$\eps =0.02$, which is in the volatile region, a broad, non-gaussian
 distribution of cycle lengths is found (see Fig.~\ref{fig:PeriodHisto}).
Furthermore, since the cycle-length
distribution shown in Fig.~\ref{fig:PeriodHisto} does not exhibit
peaks at regularly spaced intervals, the possibility that we have
employed an errant limit-cycle comparison algorithm is unlikely.
This distribution
of cycle-lengths for RSANNs differs significantly 
from the $\exp(-L^2/\tau^2)/L$ distribution of
cycle-lengths predicted for Kauffman's Boolean nets by Bastolla and Parisi
(1997). This difference in distributions implies that there may be
some significant differences between these two types of nets.

\subsection{Dependence of Attractor Count on Observation-Period Length}
\label{sec:Count}
  In Figure \ref{fig:NewCycles}, we showed that the number of
different attractors observed, ${\cal N}_{\text{cycles}}$, is a significant
fraction of the total number of trials ${\cal N}_{\text{trials}}$ when
$\eps > 0.1$. For disorder between $\eps=10^{-2}$ and $\eps=0.1$,
the number of cycles observed is much larger
than one (see Figure~\ref{fig:Div_Eps}.b), but less
than the total number of trials.  In the stable regime ($\eps <
10^{-2}$),\, ${\cal N}_{\text{cycles}}$ is quite small and largely
independent of ${\cal N}_{\text{trials}}$. These three results are
complementary: the first result (at high $\eps$)
implying a nearly inexhaustible source of
different limit-cycle attractors accompanied by a steady decrease of eligibility
with increasing $\eps$; the second result (at moderate $\eps$) implying
a large, but limited repertoire of limit cycles of high eligibility;
and the third result (at very low $\eps$)
implying that we can exhaustively access a small group
of high-eligibility limit-cycle attractors with a high degree of robustness.

 In the stable regime, ${\cal N}_{\text{cycles}}$ is often
greater than $1$ (though small), which means
that the stable phase cannot be used to access a particular attractor
upon demand, but we can demand reliable access to one of a small number of
different attractors. Frequently, however, there is a single dominant
attractor, as indicated by the low entropy seen
in Fig. \ref{fig:Stab_Entropy_Eps}.
We conclude that, in practice, for a given set of network parameters
or external inputs to the network,  but starting with different initial
conditions, the network will converge to the same attractor most of the time.

\section{Discussion}

\subsection{Ensemble-scanner, Multistability, `Creativity/Madness',
and Control Algorithms}
In the `noisy' runs at $\eps >0$, epitomized by
Fig.~\ref{fig:clock}, we are actually sampling
a significant fraction of an entire ensemble of closely-related networks
over the course of time, as the noise or disorder or an external
input slowly varies.  Based upon the results for our RSANN model,
quenched noise (disorder) can give a dynamical
system access to a whole ensemble of different behaviors at different
times during the lifetime of the dynamical system.
In other words, slowly-varying threshold
noise or disorder or external input
can act as a `scanner' for a host of dynamic modes.

The posed existence (Skarda \& Freeman (1987), Yao \& Freeman (1990),
Freeman {\it et al.} (1997)) of a chaotic ground-state attractor
for the olfactory system and the existence of `multistable' limit-cycle 
excited-state attractor lobes provides a striking exemplification of our
volatility concept, and potentially an {\it in vivo} demonstration of this
phenomenon. 

The RSANN networks studied here have a small
repertoire of behaviors when there is no noise 
or disorder in the thresholds ($\eps=0$); hence, there
is little multistability. The size of the repertoire
becomes tremendously large (as does
the extent of multistability) for `small' changes in the threshold parameters
($\eps > 10^{-2}$). One is therefore tempted to call
this behavior `chaotic' with respect to the parameter changes,
since it has one of the hallmarks of chaos (sensitive
dependence upon small changes of the parameters). However,
for smaller changes in the threshold parameters
($0 < \eps < 10^{-3}$), the repertoire of $\eps = 0$ behaviors
is `stable' -- no new cycles are observed. 
For the purpose of discussion , we refer to this
{\it delimited} sensitivity to parameter changes, as `quasi-chaos'
or `quasi-multistability'. 
Additionally, the stability is augmented by the fact that very frequently, the
repertoire of a network in the stable regime is dominated
by a single cycle (for fixed connections and fixed disorder),
a phenomenon known as `canalization' (Kauffman (1993)).
Hence, for the sake of argument, we will also 
assume that in the stable regime only one cycle is accessible.
By taking advantage of the quasi-chaotic/quasi-multistable
threshold parameters (noting that the connection
strength parameters are probably also quasi-chaotic),
we can access a large number
of different RSANN attractors, each with a small neighborhood of stability
in threshold-parameter space. With a suitable feedback algorithm,
one might be able to
control this quasi-chaos (cf. Ott, Grebogi \& Yorke (1990))
and access and stabilize a given attractor upon demand.
Due to the proximity of other limit cycles just beyond the
local neighborhood of stability of the given attractor,
 novel attractors are always
within a stone's throw of the given attractor, while 
maintaining a respectable distance so as not to be destabilizing.
Such an approach to controlled creativity
has been developed into the adaptive resonance formalism
(Carpenter \& Grossberg (1987)).

The volatile regime within $\eps=0.001-0.5$ can be subdivided into two
sub-regimes. The lower end of the range, $\eps=0.001-0.1$, which
corresponds to the upward-sloping part of the volatility curve in
Figure~\ref{fig:Vol_Eps}.b, could suggestively be named the `creative'
regime, wherein new cycles are observed with some rarity, so as to 
provide truly new behavioral modes for the RSANN. 
These new cycles can be taken together
with the more commonly-observed cycles in the net's repertoire,
perhaps to produce new and `interesting'
sequences of behavior (if these cyclic modes
can be logically sequenced). The upper end of the range,
$\eps=0.1-0.5$, might be regarded as the `overly-creative' or `slightly-mad'
regime of the RSANN. New, rather complex modes are being
found with almost every trial, which could overwhelm
the `bookkeeping' resources necessary for the RSANN to implement
or utilize the new mode to its full potential.
Clearly, this abstract and simple RSANN
model is insufficent to be a true neurobiological model
of creativity/madness, but it could be a good starting point
for a more detailed model.

\subsection{Generalization to other Complex Systems}

It may well be a common feature
of a broad class of complex, non-linear systems without adaptability or noise
that the diversity of non-trivial behaviors is limited.
We have confirmed the lack of diverse behavior
for a non-linear system with truly simple elements (McCullough-Pitts neurons);
we have also seen similar non-volatile behavior for a slightly 
more general, discretized integrate-and-fire neuron model.
This canalization result may be generalizable to other
complex systems of either simple or complex units. Indeed,
one of the first observations of canalization was in Kauffman's Boolean
immunological networks,
which have some significant differences from RSANNs.
The canalization property might have been 
more difficult to generalize if we had started with more complex units
like Hodgkin-Huxley neurons. Furthermore, it is plausible that
the introduction of a moderate amount of noise or disorder
will {\it generally} increase the diversity of complex
behaviors, as we have seen in RSANNs. 

Our volatility-producing model might be applicable in more abstract situations.
One might imagine that the states of our simple Boolean neuron reflect
in some manner the `on or off' state of complex subunits of a modular
system. Such modular complex systems 
could be probed to determine whether dynamical
diversity can or cannot be enhanced
by small changes in network parameters. Examples might include:
\begin{enumerate}
\item A random neural network composed of subnetworks;
\item A network of complex, real neurons; 
\item The brain of an organism with its different subsystems;
\item The geoeconomic or political structure of a large country composed
of smaller states, regions, or cities; and
\item The ecological network of the world composed of different regions or of 
different subcommunities of animals or plants.
\end{enumerate}
In models with subunits that are composed of many sub-subunits,
the stability or canalizing ability of the system itself may be
significantly enhanced either by a law-of-large-numbers
decrease in the noise/disorder susceptibility of an individual subunit, or
by self-stabilizing internal feedback loops which may
be present by design within the subunits.

\subsection{Conclusions \& Prospects}

Based on combinatorics and statistical arguments,
one expects to find many limit cycles in a random synchronous
asymmetric neural network (RSANN).
Experience has shown otherwise.  After much of this paper was completed,
we found an analytical argument by Amari (1974, 1989) to the effect that
RSANNs have only one attractor, in the thermodynamic limit of
a large number of neurons, thus explaining
our $\eps=0$ results.

The main objective of our study has been to construct a volatile neural
network which exhibits a large set of easily-accessible highly-eligible
limit-cycle attractors, as has been achieved already in a non-neural
system (Poon \& Grebogi (1995)). 
First, we have demonstrated
that in the absence of noise and in the absence
of random, long-term imposition of threshold disorder, a random asymmetric
neural network can reach only a small number of different
limit-cycle attractors.
Second, by imposing and freezing neuronal threshold disorder within a
well-defined range ($\eps _1<\eps <\eps _2$), we show that RSANNs can
access a diversity of highly-eligible limit-cycle attractors. RSANNs
exhibit a phase transformation from a small number of distinct
limit-cycle attractors to a large number at a disorder amplitude of
$\eps= \eps _1\sim 10^{-2}$.  Likewise, RSANNs exhibit an eligibility
phase transformation at a threshold disorder amplitude of $\eps =\eps
_2\sim 0.5$.

Potentially, Amari's argument can be extended 
to gain an understanding of how slight changes of threshold
parameters beyond some minimal level can substantially increase
the diversity of accessible cyclic modes. This extension
is beyond the scope of the current work. Another very interesting
question is how the diversity and volatility curves scale
with the size of the network.

While the addition of threshold disorder seems to be a trivial
mechanism for enhancing the volatility or diversity by constantly changing the
parameters of the RSANN, we believe that since some biological systems
(Neiman {\it et al.} (1999)) may use threshold, synapse and/or
externally-generated noise or disorder to enhance their abilities, we
have discovered a simple feature which could have some importance in
modeling biological systems.  We fully expect that other
volatility-enhancing mechanisms are available beyond the particular
one proposed here.

In summary, our key result is that a random neural network  
can be driven easily from one to another 
stable recurrent mode. While such behavior can be always
accomplished by radical modifications of some of the network properties, the
interesting result we have here presented is that plausibly small
(e.g., RMS in the neighborhood of 0.1-1\%) and random changes imposed
{\it simultaneously} upon all of the neural threshold
parameters suffices to access new dynamical behavior.
Indeed, due to the combinatorics of changing many parameters
simultaneously, an immense number of interesting modes become available
to the system. We are aware that this does not yet create
a network that can self-sequence a series of modes, though
some authors have already made considerable progress 
in this direction (e.g. Dauc\'e and Quoy (2000), Tani (1998)).
 The development of autonomous control 
algorithms that provide access to mode sequences is a natural but
challenging objective that can potentially lead
to a deeper understanding of information processing in
recurrent neural networks.

\subsection*{Acknowledgements}

H. Bohr would like to acknowledge the hospitality of
P. Carruthers (now deceased), J. Rafelski, and the
U. Arizona Department of Physics during several visits when much
of this work was carried out.  During his graduate studies, P. McGuire
was partially supported by an NSF/U.S. Department of Education/State
of Arizona doctoral fellowship.  J.W. Clark acknowledges research
support from the U.S. National Science Foundation under Grant No.
PHY-9900713. McGuire, Bohr and Clark were participants in the
Research Year on the ``The Sciences of Complexity:~From
Mathematics to Technology to a Sustainable World'' at the
Center for Interdisciplinary Studies (ZiF) at the University of Bielefeld,
in Germany.  We all thank many individuals who have provided different
perspectives to our work, including the following: G. Sonnenberg,
D. Harley, Z. Hasan, H. Ritter, R. Vilela-Mendes, and G. Littlewort.

\end{document}